\documentclass[%
 reprint,
superscriptaddress,
 amsmath,amssymb,
 aps,
]{revtex4-2}
\usepackage{braket}
\usepackage{physics}
\usepackage{amsfonts}
\usepackage{dsfont}
\usepackage{amsmath}
\usepackage{graphicx}
\usepackage{dcolumn}
\usepackage{bm}
\usepackage{xcolor}

\begin{document}


\title{Predicting three-dimensional chaotic systems with four qubit quantum systems}


\author{Joel Steinegger}
\email{{joel.steinegger@dlr.de}}
\affiliation{Deutsches Zentrum f{\"u}r Luft- und Raumfahrt (DLR), Institut f{\"u}r KI Sicherheit, Wilhelm-Runge-Stra{\ss}e 10, 89081 Ulm, Germany}
\affiliation{Deutsches Zentrum für Luft- und Raumfahrt (DLR), Institut f{\"u}r Materialphysik im Weltraum, Linder H\"ohe, 51170 K{\"o}ln, Germany}

\author{Christoph R\"ath}
\email{christoph.raeth@dlr.de (corresponding author)}
\affiliation{Deutsches Zentrum f{\"u}r Luft- und Raumfahrt (DLR), Institut f{\"u}r KI Sicherheit, Wilhelm-Runge-Stra{\ss}e 10, 89081 Ulm, Germany}

\date{January 25, 2025}

\begin{abstract}

Reservoir computing (RC) is among the most promising approaches for AI-based prediction models of complex systems. It combines superior prediction performance with very low CPU-needs
for training. Recent results demonstrated that quantum systems are also well-suited as reservoirs
in RC. Due to the exponential growth of the Hilbert space dimension obtained by increasing the
number of quantum elements small quantum systems are already sufficient for time series
prediction. Here, we demonstrate that three-dimensional systems can already well be predicted by
quantum reservoir computing with a quantum reservoir consisting of the minimal number of qubits
necessary for this task, namely four. This is achieved by optimizing the encoding of the data, using
spatial and temporal multiplexing and recently developed read-out-schemes that also involve higher
exponents of the reservoir response. We outline, test and validate our approach using eight prototypical three-dimensional chaotic systems. Both, the short-term prediction and the reproduction of
the long-term system behavior (the system's "climate") are feasible with the same setup of optimized hyperparameters. Our results may be a further step towards the realization of a dedicated small quantum computer for prediction tasks in the NISQ-era.

\end{abstract}

\maketitle


\section{\label{sec:intro}Introduction}
A fundamental challenge in various disciplines of science, engineering, medicine, and economics is the prediction of complex dynamical systems \cite{Brunton_Kutz_2019}. The ability to predict future trends and behavior from historical data could lead to many advancements in the aforementioned fields. Recent progress in the field of data-driven artificial intelligence (AI) has led to great progress in many areas, including the forecasting of complex dynamical systems \cite{machine_learning_and_chaos}. In this context reservoir computing (RC) \cite{Jaeger2001, Maass_2002, Prokhorov_2005} has emerged as a well-suited approach to predict short- and long-term properties of chaotic dynamical systems \cite{Lu_2017, Pathak_2017, Pathak_2018, Lu_2018, Parlitz_2018, Kong_2024, Li_2024} that requires only small training datasets compared to other recurrent neural networks (RNNs), does not suffer from the vanishing gradient problem, and has small computational needs. At the core of the model is a random neural network with loops called reservoir that acts as a memory and yields a reservoir state to a given input. After initialization is the network topology fixed and only the weights of a linear output layer are optimized to map the reservoir state to the right output using linear regression. This practice of linearly mapping the reservoir response results in a fast and computationally efficient training. Apart from software-based reservoirs (so-called echo state networks - ESNs)  exists the idea to realize a RC by a physical {system}, leading to novel, unconventional computers going beyond the capability of classical von Neumann computing concepts. A class of systems that are proposed for physical RC are controllable quantum systems - The exponentially large Hilbert space of a quantum system is supposed to be leveraged for time series forecasting. This branch of RC is called Quantum Reservoir Computing (QRC) \cite{Fujii_2017,Fujii_2020,Nakajima_2019, Martinez_Pena,Chen_2019, Mart_nez_Pe_a_2020, tran2020, QRC_gate_computing, Ghosh_2019, Ghosh2019QuantumNP, Chen_2020, ML_nuclear_spin_ensemble, Mujal2021OpportunitiesIQ}. QRC is motivated by advancements in the field of quantum computing and is a hybrid  classical-quantum  machine learning model. Due to the efficient, simple training, this framework is a good candidate for a quantum computing method that can outperform classical computing on NISQ era \cite{Preskill2018,NISQ_era_algorithms,Lau_2022} devices.\\
Here, we introduce a modified version of the basic QRC framework. Our novel approach combines elements from QRC already discussed in the literature (temporal \cite{Fujii_2017,Fujii_2020} and spatial multiplexing \cite{Nakajima_2019}), with state-of-the-art practices that stabilize and boost performance in ESNs, namely non-linear readout functions \cite{minimal_RC} and a proper data preprocessing pipeline. The resulting novel framework is benchmarked with prototypical synthetic chaotic systems. The focus here is two-fold. The first goal is to statistically validate the short- and long-term forecasting results of QRC. This important step is - to the best of our knowledge - so far missing in current research. Secondly, we do want to showcase that the proposed {framework} is capable of achieving very good prediction results with extremely small simulated quantum systems. This is essential for the future application on NISQ devices. In SEC. \ref{sec:results} the general task, the simulation setup and the prediction results are presented. In SEC. \ref{sec:discussion} the results are discussed in the context of future applications and open questions are addressed. Finally, in SEC. \ref{sec:methods} the model and its hyperparameter space, the details about the simulated quantum systems and the performance measures are introduced. 

\section{\label{sec:results}Results}
The modified version of the initially proposed QRC-algorithm \cite{Fujii_2017,Fujii_2020} investigated in this study is designed to forecast (continue) a $d$-dimensional discrete time series $\mathrm{\mathbf{u}}(t)=\{\mathrm{\mathbf{u}}_{j} \}_{j=1}^{L}$ $=\{\mathrm{\mathbf{u}}(t_{0}),\mathrm{\mathbf{u}}(t_{0}+\Delta t),\cdots \}$ from past time steps. Meaning, the model is supposed to approximate a function $f$ that fulfils
\begin{equation}
\label{temporal_task}
\mathrm{\mathbf{u}}_{k+1} = f (\{\mathrm{\mathbf{u}}_{j} \}_{j=1}^{k}).
\end{equation}
and generalizes for unseen data. {As mentioned in the introduction, RC approaches are well suited to solve such a task. 
Input data (in discrete time steps) are recurrently injected into the reservoir. The dynamics of the reservoir produces a high-dimensional and non-linearl reservoir response that encodes information of the current state and the recent past of the dynamical system. The readout layer is trained to map this reservoir response (or readout vector) to the next step of the time series. To this end, QRC leverages the exponentially large Hilbert space of quantum systems (here multiple qubit systems), as an enhanced feature space. This is done in a way that the quantum state retains information about the present and past of the time series. The full algorithm is defined and explained in detail in SEC. \ref{sec:algorithm}. It follows a short description of the above described recurrent process. For each step $k$ of the discrete time series, the current state of the time series is encoded into $r$ quantum systems. The systems are then evolved by a unitary operator which scrambles the information. Thereafter, preselected expectation values are measured. This is done $V$ times for each system before the next timstep $k+1$ is encoded into the systems. By collecting all these observables and also including higher exponents of these observables as additional nodes the output vector $\mathrm{\mathbf{q}}(k)$ of step $k$ (reservoir response of step $k$) is obtained. These vectors are trained by Ridge regression to linearly map the reservoir response onto the subsequent step of the time series and therefore fulfilling the above defined task.} 
\subsection{\label{sec:experimental_setup}{Simulation} setup}
In this study the prediction performance of the QRC model is investigated by forecasting eight different prototypical three-dimensional chaotic systems like the Lorenz-63 and the R\"ossler system (defined in Supplemental information S4) with a numerical simulation of small controllable quantum systems. To showcase the predictive power of the QRC model when very small qubits systems are accessible, all quantum systems are selected to be as small as the algorithm theoretically allows. Therefore all the results in this work are obtained with four qubit systems. Information regarding the simulated quantum system's unitary evolution can be found in Sec. \ref{sec:unitary_evolution}. The time series are preprocessed by standardization and subsequently scaled into the interval [$a$,$b$] with $0\le a<b\le1$. This practice gives rise to two new hyperparameters: $a$ and $b$. The measured expectation values that are selected to built the output vectors $\mathrm{\mathbf{q}(k)}$ are the spin-projection $\langle \sigma_{\mathrm{z}}^{i} \rangle$ and the spin-correlation $\langle \sigma_{\mathrm{z}}^{i}\sigma_{\mathrm{z}}^{l} \rangle$ with $i,l \in \{1,\hdots,4\} \ \text{and} \ i<l$.
The model has some free hyperparameters. The parameters that are part of the classical part of the algorithm are the scaling parameters $a$ and $b$, the regression parameter $\mathrm{\beta}$ and the degree $G$ of the used exponents of the reservoir readout. The free hyperparameters of this study concerning the quantum part of the algorithm are the number $r$ of employed quantum systems and the number of employed evolution and measurement processes $V$ per encoded time step $k$.\\
The results in this study are obtained for each investigated hyperparameter combination by evaluating the model for $N_{\mathrm{stat}}$ times. For each of these runs, different parts of the chaotic attractors and different choices of the random parameters for the unitary transformations describing the quantum systems are selected to enable statistical significant performance evaluation. Each model is trained (see Sec. \ref{sec:algorithm}) with $N_{\mathrm{sync}}+N_{\mathrm{train}}=L$ consecutive steps of the trajectory and subsequently the time series is continued for $N_{\mathrm{pred}}$ steps.

\subsection{\label{sec:prediction_results}Prediction results}
\begin{figure*}[htb]
    \begin{center}
    \includegraphics[width=\textwidth]{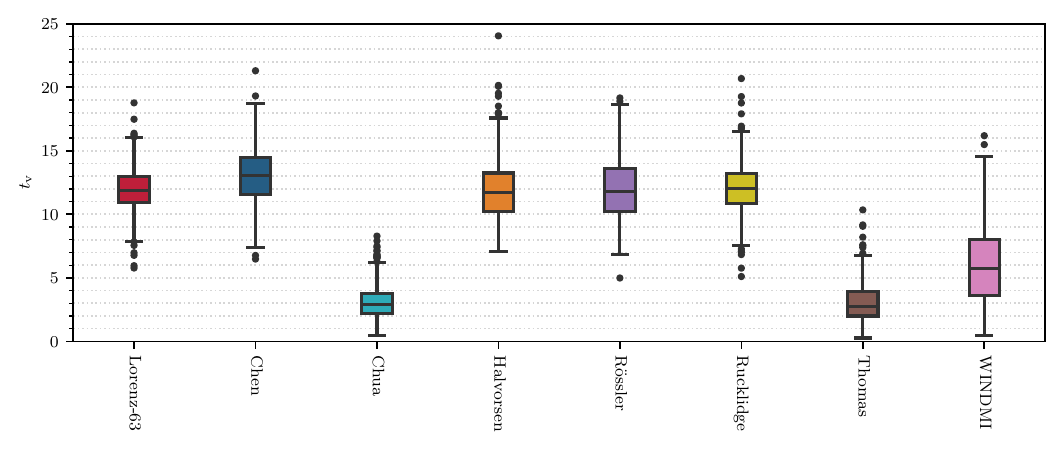}
    \end{center}
    \caption{\label{fig:forecast_horizon}Distributions of the short-term accuracy for all eight forecasted chaotic systems. These are shown in a boxplot of the forecast horizon in Lyapunov times for the dynamical systems. The boxes represent the 25$\%$-75$\%$ percentile range of the data and the line in the middle of the box shows the median of the data, i.e. 50$\%$ of the forecast horizons are below this value. The extended lines showcase largest and smallest observation that falls within a distance of 1.5 times the interquartile distance (IQR) of the data. The black dots represent the outliers that are outside of this range. The mean and standard deviations of the distributions can be found in Table \ref{table:performance}.}
\end{figure*}

\begin{figure*}[htb]
    \begin{center}
    \includegraphics[width=\textwidth]{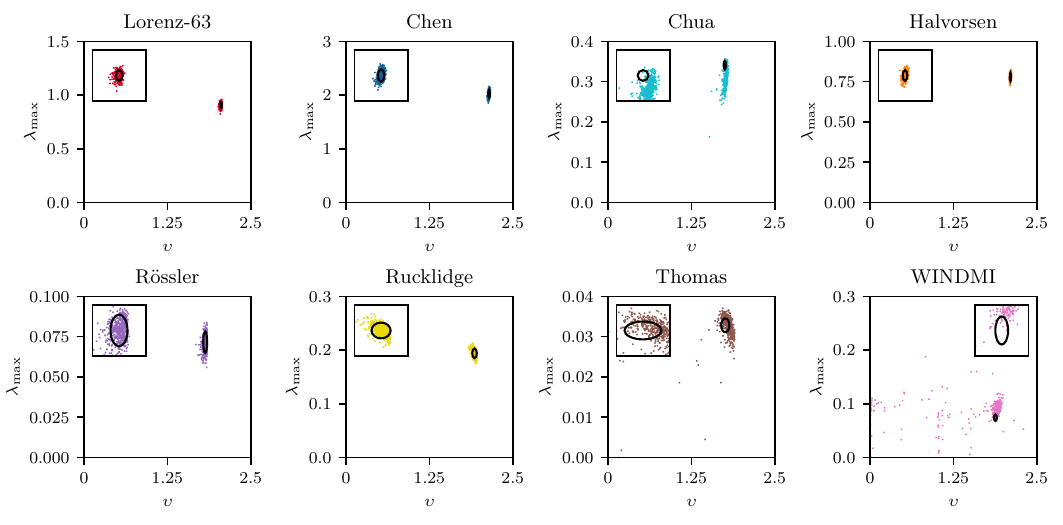}
    \end{center}
    \caption{\label{fig:climate}Distributions of the predicted climate for all eight forecasted chaotic systems. These are depicted by showing the Largest Lyapunov exponent vs correlation dimension for each of the 500 forecasted trajectories for all eight chaotic systems. The back ellipses show the three standard deviation errors of the correlation dimension and the largest Lyapunov exponent calculated from simulations of the actual systems. The zoomed-in windows plotted as insets are centered at $x= <\upsilon>_{true}$, $y = <\lambda_{\mathrm{max}}>_{true}$) and extends $\pm 15 \%  <\lambda_{\mathrm{max}}>_{true}$ in the y-direction and $\pm 5\%  <\upsilon>_{true}$ in the x-direction. The mean and standard deviations of the distributions can be found in Table \ref{table:performance}.}
\end{figure*}

In this work, we show that the introduced and simulated QRC model is capable of achieving short-term forecasting quality rivaling and in some cases outperforming "classical" RC methods and also accurately predicting the long-term climate of three-dimensional chaotic systems while utilizing four qubit systems and small training datasets $(N_{\mathrm{sync}}=100$ and $N_{\mathrm{train}}=2000$). Our main goal is to introduce techniques that make real world QRC-applications on near-term available quantum computers realistic. We choose a quantum system (unitary evolution) that works (more details SEC. \ref{sec:unitary_evolution}) and do not further optimize it. The selected hyperparameters critically influence the prediction performance in a non-trivial way (see Supplemental information S1). Covering the hyperparameter space with a fine grid search to find the best performing hyperparameter combination is out of reach due to computational limitations. Instead, a Bayesian hyperparameter optimization using the python package Optuna \cite{baysian_optimization_optuna} over a hyperparameter space section is applied. For all eight investigated chaotic systems, the best performing configuration of hyperparameters is obtained by maximizing the mean forecast horizon (defined in Sec. \ref{sec:measures}) of the model ($N_{\mathrm{stat}}$=100). With these best performing hyperparameter sets we train the model $N_{\mathrm{stat}}$=500 times and forecast the trajectories for $N_{\mathrm{pred}}$=20000 steps to evaluate the short- and long-term prediction efficiency of the model. {The short-term prediction efficiency is measured by the forecast horizon, which determines the time for which the model prediction matches the actual continuation of the forecasted dynamical system up to a small deviation. The long-term prediction efficiency is determined by how well the model is able to recreate statistical properties of the attractor of the dynamical systems. Here we use as measures the largest Lyapunov exponent and correlation dimension.} All the measures used to evaluate the prediction quality of these best performing models are defined {in detail} in Sec. \ref{sec:measures}. The inspected hyperparameter section and the best performing hyperparameter set for each chaotic system can be found in the Supplemental information (S2).\\ 
In many applications, it is important that the machine learning model is able to forecast the time series of a dynamical system very accurately for as long as possible. The best performing hyperparameter configurations are obtained by maximizing this ability. The mean forecast horizon of the 500 trained models can be found for each of the chaotic systems in Table \ref{table:performance} and the distributions are displayed as a boxplot in Fig.\ref{fig:forecast_horizon}. A comparison of our results with those of \cite{Hybrid_RC} shows that the mean forecast horizon is in all cases at least comparably long as in the classical RC approach. {Yet in some cases the QRC-models even outperform (larger mean forecast horizon) some hybrid RC approaches. This means that these simulated QRC models which are purely data-driven are able to forecast chaotic systems on longer time scales more accurately than some methods making use of prior knowledge about the physics of the underlying equations of the chaotic systems. This is especially true when the reservoir size is small in conventional hybrid RC.} Investigating Table \ref{table:performance} and Fig. \ref{fig:forecast_horizon} shows that three (Chua, Thomas and WINDMI) of the eight chaotic systems are predicted accurately on a much shorter timescale than the other ones. Interestingly enough, these systems are also comparatively badly predicted with conventional RC. It is certainly an obvious and important question to figure out the causes for the systematic differences in performance among the model systems. Yet, this research topic is beyond the scope of this paper.\\  
In some other application scenarios, the focus might not be on the short-term behavior of the system, but rather on whether a dynamical system's long-term properties (its "climate") can be reproduced correctly. We investigate this by calculating two measures quantifying the (strange) attractors of the dynamical systems, namely the correlation dimension and the largest Lyapunov exponent (both defined in Sec. \ref{sec:measures}). For every chaotic system we use 500 trajectories, all consisting of 20000 steps that were not taken for training the models, to calculate the mean largest Lyapunov exponent and correlation dimension of the systems from true trajectories. We compare the calculated statistics of the attractors with the statistics of the forecasted time series. These findings are presented in  Fig. \ref{fig:climate} and the mean values are listed in Table \ref{table:performance}. The spatial and temporal statistical properties of the five chaotic systems that are forecasted accurately for long time scales are extremely well predicted. 
In our sample, there are no single outliers with large deviations for the Lyapuov exponent or the correlation dimension. Rather, all realizations are within a $\approx 5 \sigma$ error range of the two measured quantities.
For the remaining three systems, the climate of the systems is reproduced well in some cases, but some forecasted trajectories exhibit long-term behavior that is far from the statistical fluctuations of the true data. 
To the best of our knowledge, it is shown for the first time that QRC is capable to also reproduce the statistical long-term properties of predicted chaotic time series. These findings suggest that our (minimal) QRC setup does not only learn patters of the time series by heart leading to good short term predictions but rather correctly learns the dynamics of the underlying chaotic systems, enabling correct long term predictions. 


\begin{table*}
    \centering
    \resizebox{\textwidth}{!}{
    \begin{tabular}{|l|c|c|c|c|c|}
    \hline
         system& mean $t_{v}$ & predicted mean $\lambda_{\mathrm{max}}$ & true mean $\lambda_{\mathrm{max}}$ & predicted mean $\upsilon$&  true mean $\upsilon$\\
    \hline
         Lorenz-63& 11.9 $\pm$ 1.7 & 0.91 $\pm$ 0.02 &0.91 $\pm$ 0.02 & 2.053 $\pm$ 0.008& 2.052 $\pm$ 0.009\\
         Chen &13.0 $\pm$ 2.2 & 2.02 $\pm$ 0.05 & 2.02$\pm$ 0.05 & 2.146 $\pm$ 0.008& 2.145$\pm$ 0.008\\
         Chua& 3.1 $\pm$ 1.4 & 0.31 $\pm$ 0.02 & 0.341$\pm$ 0.007 & 1.77 $\pm$ 0.02&  1.75$\pm$ 0.01\\
         Halvorsen& 11.9$\pm$ 2.3 & 0.78 $\pm$ 0.02 & 0.78 $\pm$ 0.02 & 2.106 $\pm$ 0.006&  2.106 $\pm$ 0.006\\
         Rössler&11.9 $\pm$ 2.4 & 0.071 $\pm$ 0.005 & 0.072 $\pm$ 0.004 & 1.82 $\pm$ 0.02&  1.82 $\pm$ 0.02 \\
         Rucklidge& 12.1$\pm$ 2.0 & 0.194 $\pm$ 0.006 & 0.194$\pm$0.006 & 1.93 $\pm$ 0.02& 1.93$\pm$ 0.02\\
         Thomas&3.0 $\pm$ 1.5 & 0.03 $\pm$ 0.004 & 0.033 $\pm$ 0.001 & 1.8 $\pm$ 0.2& 1.76 $\pm$ 0.04\\
         WINDMI&5.9 $\pm$ 2.9 & 0.08$\pm$ 0.04 & 0.074$\pm$ 0.004 & 1.5 $\pm$ 0.8 & 1.88$\pm$ 0.02\\
    \hline
    \end{tabular}}
    \caption{\label{table:performance}Table of mean values with standard deviation as derived from 500 realizations of all three evaluation measures for the chosen hyperparameter configuration of all eight chaotic systems. For comparison the results of the largest Lyapunov exponent and the correlation dimension for true trajectories are listed as well.}
\end{table*}
\section{\label{sec:discussion}Discussion and Conclusions}
In this work, we present a QRC architecture that is suitable for small quantum systems and performs very well in forecasting low-dimensional chaotic dynamical systems. The eight systems can be predicted on at least a few Lyapunov times very accurately. We demonstrated for the first time that the model is also able to recreate the long term dynamics of the chaotic systems in most cases very well. A future goal should be to decrease the number and optimally fully remove those realizations in the three badly performing systems that completely diverge. First steps for such a procedure are sketched in the Supplemental information (S3). We find indications that the performance of the short term predictions and the correct reproduction of the long term properties of the system are related to the actual choice of random variables controlling the spin-spin interactions and the onsite disorder in the Ising model of the quantum reservoir. We want to highlight the fact that using multiple quantum systems in combination with multiple evolution and measurement processes is a key to achieve {optimal} performance with small quantum systems. All best performing hyperparameter configurations use the maximal number of reservoirs ($r=3$) of the range we selected for the Bayesian hyperparameter optimization. Furthermore, finding a good regression parameter is important for decreasing badly performing outliers. Our investigation of the hyperparameters a,b is very simplistic. Whether the performance can be increased by optimizing the hyperparameters over the continuous interval $0\le a<b\le1$ is an open question. {Other choices of how the time series is encoded might also help to achieve better forecasting ability of the models.} Another open question is whether the performance can be further improved by increasing $G$ or by using different readout functions $\mathrm{F_{res}}$. Other methods that are used to increase prediction performance in "classical" RC like adding noise to the training data to decrease overfitting and using other regression algorithms (e.g. tree-regression \cite{giammarese2024}) should also be investigated.\\ 
In conclusion, the results of our research in combination with more suitable unitary evolution for NISQ-devices could put hardware realizations and real-world applications of QRC not far out of reach. 
Future work should further investigate the employed unitary operator and look further into  what defines a good performing quantum reservoir while ideally keeping NISQ-device restrictions {(e.g. noise)} in the hyperparameter and unitary operator selection in mind. 

\section{\label{sec:methods}Methods}
\subsection{\label{sec:algorithm}QRC framework}
\begin{figure}[h!]
    \begin{center}
    \includegraphics[width=\linewidth]{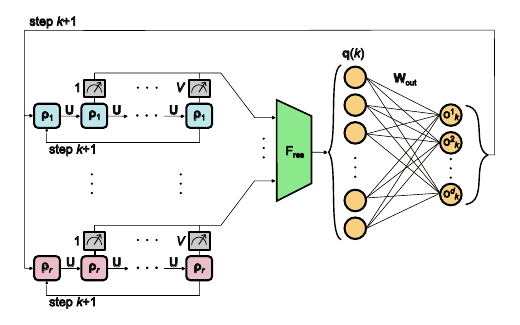}
    \end{center}
    \caption{\label{fig:Prediction} Schematic illustration of the prediction phase of the QRC model.}
\end{figure}

The input data is sequentially encoded for each step of the time series into the quantum system. For an $N$ qubit system, this sequential input is realized by successively initializing the quantum state for each step of the time series into the state
\begin{equation}
\label{encoding}
\begin{tiny}
{\rho_{k} = \rho_{{\mathrm{\mathbf{u}^{1}}_{k}}} \otimes ...\otimes \rho_{{\mathrm{\mathbf{u}}^{d}_{k}}} \otimes \mathrm{Tr}_{1,\cdots,d}(\rho(k-1)))}
\end{tiny}
\end{equation}
where $\rho_{{\mathrm{u}^{i}_{k}}}=\ketbra{{\mathrm{u}^{i}_{k}}}{{\mathrm{u}^{i}_{k}}}$ and $\ket{{\mathrm{u}^{i}_{k}}}=\sqrt{{1-\mathrm{ u}^{i}_{k}}}\ket{0}+\sqrt{{\mathrm{u}^{i}_{k}}}\ket{1}.$
Here and in the following, the $i$-th coefficient of the $k$-th step of the discrete time series $\mathrm{\mathbf{u}}(t)$ is denoted as $\mathrm{u}^{i}_{k}$ and $\mathrm{Tr}_{1,\hdots,d}(.)$ is the partial trace over the first $d$ qubits. This encoding choice has two obvious consequences. The first one is that to retain information about past inputs, the number of qubits has to be larger than the dimension of the time series. The second one is that the time series has to be scaled to the interval [0,1]. Following the encoding of one time step the system evolves under unitary evolution. The full map between two quantum states of the sequence is
\begin{equation}
\begin{tiny}
\label{QRC_evolve}
\rho(k) = \mathrm{U} {\rho_{k}} \mathrm{U}^{\dag}.
\end{tiny}
\end{equation}
Following the unitary evolution, expectation values are used to extract the high dimensional encoded information of the present and recent history of the time series. For each step of the time series, these expectation values (output nodes) are collected in the reservoir output vector $\mathrm{\mathbf{n}}(k)$ of step $k$. The dimensionality of this vector is determined by the choice of observables and the number of qubits of the quantum system. One of the primary aims of this paper is to showcase the capacity QRC has even when only very small quantum systems are available. To increase the number of output nodes without increasing the number of qubits there are three methods employed.\\
\underline{Temporal multiplexing} \cite{Fujii_2017,Fujii_2020}: Rather than employing the unitary evolution and the measurement one time before encoding the next step of the time series, the evolution and measurement process is carried out $V$ times. The output vectors for each of these single measurement phases are merged together into one output vector 
\begin{equation}
\label{Output_layer_virtual_nodes}
\begin{tiny}
\mathrm{\mathbf{v}}(k)=\begin{pmatrix} \mathrm{\mathbf{n}}_{1}(k)  \\ \vdots\\ \mathrm{\mathbf{n}}_{V}(k)\end{pmatrix}
\end{tiny}
\end{equation}
of step $k$.\\
\underline{Spatial multiplexing} \cite{Nakajima_2019}: Rather than employing just one quantum system, multiple reservoirs are used, and the output vector of each of these reservoirs are concatenated together into one output vector of step $k$
\begin{equation}
\label{Output_layer_spartial}
\begin{tiny}
\mathrm{\mathbf{p}}(k)=\begin{pmatrix}  \mathrm{\mathbf{v}}_{1}(k) \\  \vdots\\ \mathrm{\mathbf{v}}_{r}(k)  \end{pmatrix}.
\end{tiny}
\end{equation}
\underline{Reservoir readout function:}
The final technique to increase the dimension of the output vector is to apply a function $\mathrm{F_{res}}$. In this work, powers of the reservoir readout up to the fourth order are considered as the readout function. The studied choices are thus only the reservoir response (and a bias term)
\begin{equation}
\label{final_output_vector_identity}
\begin{tiny}
\mathrm{\mathbf{q}}(k)=\mathrm{F_{res}}( \mathrm{\mathbf{p}}(k))=\mathrm{\begin{pmatrix}  1 \\\mathrm{\mathbf{p}}(k)   \end{pmatrix}},
\end{tiny}
\end{equation}
the reservoir response and its square (also known as "Lu readout"\cite{Lu_2017})
\begin{equation}
\label{final_output_vector_Lou}
\begin{tiny}
\mathrm{\mathbf{q}}(k)=\mathrm{F_{res}}(\mathrm{\mathbf{p}}(k))=\begin{pmatrix}  1 \\ \mathrm{\mathbf{p}}(k) \\\mathrm{\mathbf{p}}^{2}(k) \end{pmatrix},
\end{tiny}
\end{equation}
the reservoir response and its second and third power
\begin{equation}
\label{final_output_vector_cubed}
\begin{tiny}
\mathrm{\mathbf{q}}(k)=\mathrm{F_{res}}(\mathrm{\mathbf{p}}(k))=\begin{pmatrix}  1 \\ \mathrm{\mathbf{p}}(k)\\\mathrm{\mathbf{p}}^{2}(k) \\\mathrm{\mathbf{p}}^{3}(k)  \end{pmatrix},
\end{tiny}
\end{equation}
and the reservoir response and its powers up to four
\begin{equation}
\label{final_output_vector_quartic}
\begin{tiny}
\mathrm{\mathbf{q}}(k)=\mathrm{F_{res}}(\mathrm{\mathbf{p}}(k))=\begin{pmatrix}  1 \\ \mathrm{\mathbf{p}}(k)\\ \mathrm{\mathbf{p}^{2}}(k) \\ \mathrm{\mathbf{p}}^{3}(k)\\ \mathrm{\mathbf{p}}^{4}(k)\end{pmatrix}.
\end{tiny}
\end{equation}
This selection of readout functions adds the hyperparameter $G$ indicating the powers of the readout being included in the model. This method is inspired by new findings in "classical" RC, where it was shown that shifting the nonlinearities to the readout layer yields good prediction results even for minimal reservoirs \cite{minimal_RC}.\\ 
Hereafter, it is explained how these output vectors are exploited for time series forecasting.\\
\textbf{Training}: The quantum systems are initialized in the quantum state
\begin{equation}
\label{start state}
\mathrm{\rho(0)=(\ket{0}\bra{0})}^{\otimes N}.
\end{equation}
For a given training data set $\{\mathrm{\mathbf{u}}_{j} \} _{j=1}^{L} $ the first $N_{\mathrm{sync}}$ steps are used to synchronize the quantum systems with the dynamics of the input data and thereby eliminate any transient dynamics that are a product of the initialization of the quantum states. The synchronization is accomplished by injecting the time series sequentially into the quantum reservoirs and letting the quantum states evolve as previously defined. The leftover $N_{\mathrm{train}}$ steps of the training data set are used to obtain a readout matrix $\mathrm{\mathbf{W}_{out}}$ such that the quantum reservoir output vector $\mathrm{\mathbf{q}}(k)$ is mapped to step $k+1$ of time series $\mathrm{\mathbf{u}}_{k+1}$. To do this, the system is sequentially injected with the remaining training data. The full output vectors $\mathrm{\mathbf{q}}(k)$ are measured and collected in a matrix $\mathrm{\mathbf{Q}}=[\mathrm{\mathbf{q}}(N_{\mathrm{sync}}+1),\hdots,\mathrm{\mathbf{q}}(N_{\mathrm{sync}}+N_{\mathrm{train}}-1) ]$ and the desired outputs are collected in a matrix $\mathrm{\mathbf{Y}}=[\mathrm{\mathbf{u}}_{N_{\mathrm{sync}}+2},\hdots,\mathrm{\mathbf{u}}_{N_{\mathrm{sync}}+N_{\mathrm{train}}}]$. The objective is to find a matrix $\mathrm{\mathbf{W}_{out}}$ that solves the equation
\begin{equation}
\label{QRC_Train_equation}
\mathrm{\mathbf{Y}=\mathbf{W}_{out}\mathbf{Q}}.
\end{equation}
Ridge regression (detailed in Sec. \ref{sec:ridge_regression}) is used to obtain $\mathrm{\mathbf{W}_{out}}$. 
\noindent\\
\textbf{Prediction}: 
Once the matrix $\mathrm{\mathbf{W}_{out}}$ is obtained, arbitrary long predictions continuing the time series can be calculated. To continue the time series, an autonomously evolving closed-loop is employed. This is achieved by using the last prediction of the model
\begin{equation}
\label{Prediction_qrc}
\mathrm{\mathbf{o}}_{k+1}=\mathrm{\mathbf{W}_{out}\mathbf{q}}(k) \ \text{with} \ k>L-1.
\end{equation}
as the next input. The optimized readout matrix is kept fixed throughout the whole forecasting process. The prediction phase is schematically illustrated in Fig. \ref{fig:Prediction}. The continued time series is denoted as $\mathrm{\mathbf{y}_{pred}}(t)=\{\mathrm{\mathbf{o}}_{k}\}_{k=L+1}^{L+N_{\mathrm{pred}}}$ in the following.
\subsection{\label{sec:ridge_regression}Ridge Regression}
Ridge regression is used to obtain the readout matrix $\mathrm{\mathbf{W}_{out}}$ by calculating
\begin{equation}
\label{QRC_Ridge_regression_Wout}
\mathrm{ \mathbf{W}_{out}=\mathbf{Y}\mathbf{Q}^{T}(\mathbf{Q}\mathbf{Q}^{T}-\beta \mathds{1})^{-1}}
\end{equation}
which minimizes the error function
\begin{equation}
\label{Ridge_regression_error}
\sum_{N_{\mathrm{sync}}<k<N_{\mathrm{train}}} \norm{\mathrm{\mathbf{W}_{out}\mathbf{q}}(k)-\mathrm{\mathbf{u}}_{k}}^{2}+\beta \mathrm{Tr(\mathbf{W}_{out}^{T}\mathbf{W}_{out})}.
\end{equation}
$\mathrm{\beta}$ is an important hyperparameter that improves the prediction by penalizing large matrix coefficients.

\subsection{\label{sec:unitary_evolution}Simulation details: Unitary operator}
In this work, a spin network described by the transverse field Ising model plus onsite disorder 
\begin{equation}
\label{hamiltonian_TFIM}
\mathrm{H} = \sum_{i>j=1}^{N}J_{ij}\sigma_{\mathrm{x}}^{i}\sigma_{\mathrm{x}}^{j}+\frac{1}{2}\sum_{i=1}^{N}(h+D_{i})\sigma_{\mathrm{z}}^{i}
\end{equation}
is chosen as the reservoir in alignment with previous QRC research \cite{Martinez_Pena}. In the equation above, $N$ expresses the number of qubits and $h$ is the magnetic field. The spin-spin couplings $J_{ij}$ are randomly selected from a uniform distribution in the interval [$-J/2, J/2$]. In the same manner, the onsite-disorders $D_{i}$ are randomly selected from the interval [$-W,W$]. Finally, $\sigma_{a}^{i}$ with $a \in$ $\{$x,y,z$\}$ denote the Pauli-matrices. The time evolution of the quantum system is utilized as the unitary operator. A unit time step size $\tau$ is chosen as the time between two consecutive inputs, and the observable measurements are carried out $V$ times after letting the reservoirs evolve for a time $\tau/V$. The unitary operator that maps between states is
\begin{equation}
\label{unitary_evolution_hamiltonian}
\mathrm{U} = e^{\frac{-i\mathrm{H}\tau}{V}}.
\end{equation}
Motivated by previous results \cite{Martinez_Pena} the unit time step size $\tau=20J$ is not optimized and the quantum systems are chosen to be in the ergodic phase with $h=2/J$ and $W = 0.05/J$. All parameters are expressed in units of $J$. For convenience, $J$=1 is selected in the simulations. We described in Sec. \ref{sec:algorithm} the model as general as possible because the choice of unitary operator is not unique. The appropriate unitary operator is going to be dependent on constraints of the available NISQ-devices. In \cite{QRC_gate_computing} unitary operator choices are investigated that might be more suitable for applications in the near-term future. The model introduced in our work is applicable to other unitary operators and therefore gives a good starting point for further research towards employing small quantum system for time series forecasting. 

\subsection{\label{sec:measures}Prediction performance measures}
To evaluate the quality of the predictions, three different measures are used. These measures are chosen to sufficiently assess the quality of the exact short-term prediction and the reproduction of the long-term statistical properties (climate) of the systems. The measures used for the evaluation follow previous studies \cite{RC_1,RC_2,RC_3} investigating "classical" RC.\\

The forecast horizon (also called valid time) is calculated to evaluate the short-term prediction capabilities of the model. It measures the time for which the continued time series matches very closely the true continuation of the trajectories. The forecast horizon is the elapsed time, while the normalized, time-dependent error $e(t)$ \cite{forecast_horizon} between the continued time series $\mathrm{\mathbf{y}_{pred}}(t)$ and the true continuation $\mathrm{\mathbf{y}}(t)=\{\mathrm{\mathbf{u}}_{k}\}_{k=L+1}^{L+N_{\mathrm{pred}}}$ is smaller than a threshold value $e_{\mathrm{max}}$. The normalized, time-dependent error is defined as 
\begin{equation}
\label{time_dependent_error}
e(t)=\frac{\Vert \mathrm{\mathbf{y}}(t) - \mathrm{\mathbf{y}_{pred}}(t) \|}{\langle\Vert  \mathrm{\mathbf{y}}(t)  \|^{2}\rangle^{1/2}}.
\end{equation}
Here, $\mathrm{\langle . \rangle}$ denotes the average over all $N_{\mathrm{pred}}$ steps and $\Vert . \|$ is the L2-norm. It is determined for how many consecutive steps $s_{v}$ (starting with the first forecasted state) the relation $e(t)<e_{\mathrm{max}}$ holds. In this work the threshold is chosen to be $e_{\mathrm{max}}$=0.4. The forecast horizon $t_{v}$, in units of the Lyapunov times 1/$\lambda_{\mathrm{max}}$ of the dynamical system, is obtained by calculating 
\begin{equation}
\label{forecast_horizon_calc}   
t_{v}=\Delta t	\cdot s_{v} \cdot \lambda_{\mathrm{max}}.
\end{equation}
Here, $\Delta t$ is the time between two successive steps of the discretized time series, and $\lambda_{\mathrm{max}}$ is the largest Lyapunov exponent (defined in the following paragraphs) of the system. The forecast horizon is measured in Lyapunov times to obtain a measure that is more comparable across different dynamical systems.\\
One aspect of the long-term behavior of a dynamical system is its structural complexity. The correlation dimension is a measure that quantifies the structural complexity by measuring the dimensionality of the space populated by the trajectory \cite{GRASSBERGER_correlation_dimension}. This measure is based on the discrete version of the correlation integral 
\begin{equation}
\label{eq: correlation_integral}
C(r)=\lim \limits_{M \to \infty} \frac{1}{M^{2}} \sum_{i,j=1}^{M} \Theta (r - \Vert \mathrm{\mathbf{x}}_{i}-\mathrm{\mathbf{x}}_{j} \|).
\end{equation}
Here, $\Theta$ represents the Heaviside function. $C(r)$ calculates the mean probability that two states in phase space are closer than a threshold distance $r$ for different time steps. For a self-similar strange attractor, the correlation dimension is defined by a power-law relationship in a certain range of the threshold $r$:
\begin{equation}
\label{eq: correlation_integral_relationship}
C(r)\propto r^{\upsilon}.
\end{equation}
The scaling exponent $\upsilon$ gives the correlation dimension of the attractor. The Grassberger Procasccia algorithm \cite{Grassberger_Procaccia} is used to calculate the correlation dimension from data.\\
Another characteristic of the long-term climate of a system is its temporal complexity. The most appropriate way to quantify the temporal complexity of a dynamical system is to analyze its Lyapunov exponents, characterizing the system's development in time. A $d$-dimensional dynamical system has $d$ Lyapunov exponents that determine the average rate of divergence of nearby points in phase space. By measuring the average rate of exponential growth of a small perturbation in each direction in phase space, the Lyapunov spectrum measures how sensitive the system is to its initial conditions. A dynamical system exhibits chaos if one of its Lyapunov exponents is positive, and the magnitude of the exponent determines the timescale on which the system becomes unpredictable \cite{Wolf_1985,Shaw}. The largest Lyapunov exponent $\lambda_{\mathrm{max}}$ is linked to the direction in which the divergence occurs most rapidly,
\begin{equation}
\label{eq: lambda_max}
d(t)=c \cdot e^{\lambda_{\mathrm{max}}t}.
\end{equation}
In this research, measuring the largest Lyapunov exponent suffices, because of its dominant influence over the dynamics. This constraint also holds a computational advantage because $\lambda_{\mathrm{max}}$ can be easily calculated from data using the Rosenstein algorithm \cite{Rosenstein_algorithm}.
\section{Data availability}
The data that support the findings of this study are available from the corresponding 
author upon reasonable request.
\section{Code availability}
The code used to generate the findings of this study is available from the corresponding 
author upon reasonable request.
\section{Acknowledgments}
We wish to acknowledge useful discussions and comments from Fabian Fischbach, Markus Gross, Sebastian Baur and Andreas Spörl. This project was made possible by the DLR Quantum Computing Initiative and the Federal Ministry for Economic Affairs and Climate Action; qci.dlr.de/projects/nemoqc.

\section{Author contributions}
C.R. initiated and supervised the research. J.S. performed the computation and analyzed the data. J.S. and C.R. interpreted the results and wrote and edited the manuscript.
F.K. and C.R. acquired funding for the project.
\section{Competing interests}
The authors declare to have no competing interests.


\begin{thebibliography}{0}%
\makeatletter
\providecommand \@ifxundefined [1]{%
 \@ifx{#1\undefined}
}%
\providecommand \@ifnum [1]{%
 \ifnum #1\expandafter \@firstoftwo
 \else \expandafter \@secondoftwo
 \fi
}%
\providecommand \@ifx [1]{%
 \ifx #1\expandafter \@firstoftwo
 \else \expandafter \@secondoftwo
 \fi
}%
\providecommand \natexlab [1]{#1}%
\providecommand \enquote  [1]{``#1''}%
\providecommand \bibnamefont  [1]{#1}%
\providecommand \bibfnamefont [1]{#1}%
\providecommand \citenamefont [1]{#1}%
\providecommand \href@noop [0]{\@secondoftwo}%
\providecommand \href [0]{\begingroup \@sanitize@url \@href}%
\providecommand \@href[1]{\@@startlink{#1}\@@href}%
\providecommand \@@href[1]{\endgroup#1\@@endlink}%
\providecommand \@sanitize@url [0]{\catcode `\\12\catcode `\$12\catcode
  `\&12\catcode `\#12\catcode `\^12\catcode `\_12\catcode `\%12\relax}%
\providecommand \@@startlink[1]{}%
\providecommand \@@endlink[0]{}%
\providecommand \url  [0]{\begingroup\@sanitize@url \@url }%
\providecommand \@url [1]{\endgroup\@href {#1}{\urlprefix }}%
\providecommand \urlprefix  [0]{URL }%
\providecommand \Eprint [0]{\href }%
\providecommand \doibase [0]{https://doi.org/}%
\providecommand \selectlanguage [0]{\@gobble}%
\providecommand \bibinfo  [0]{\@secondoftwo}%
\providecommand \bibfield  [0]{\@secondoftwo}%
\providecommand \translation [1]{[#1]}%
\providecommand \BibitemOpen [0]{}%
\providecommand \bibitemStop [0]{}%
\providecommand \bibitemNoStop [0]{.\EOS\space}%
\providecommand \EOS [0]{\spacefactor3000\relax}%
\providecommand \BibitemShut  [1]{\csname bibitem#1\endcsname}%
\let\auto@bib@innerbib\@empty
\end{thebibliography}%


\begin{thebibliography}{41}%
	\makeatletter
	\providecommand \@ifxundefined [1]{%
		\@ifx{#1\undefined}
	}%
	\providecommand \@ifnum [1]{%
		\ifnum #1\expandafter \@firstoftwo
		\else \expandafter \@secondoftwo
		\fi
	}%
	\providecommand \@ifx [1]{%
		\ifx #1\expandafter \@firstoftwo
		\else \expandafter \@secondoftwo
		\fi
	}%
	\providecommand \natexlab [1]{#1}%
	\providecommand \enquote  [1]{``#1''}%
	\providecommand \bibnamefont  [1]{#1}%
	\providecommand \bibfnamefont [1]{#1}%
	\providecommand \citenamefont [1]{#1}%
	\providecommand \href@noop [0]{\@secondoftwo}%
	\providecommand \href [0]{\begingroup \@sanitize@url \@href}%
	\providecommand \@href[1]{\@@startlink{#1}\@@href}%
	\providecommand \@@href[1]{\endgroup#1\@@endlink}%
	\providecommand \@sanitize@url [0]{\catcode `\\12\catcode `\$12\catcode
		`\&12\catcode `\#12\catcode `\^12\catcode `\_12\catcode `\%12\relax}%
	\providecommand \@@startlink[1]{}%
	\providecommand \@@endlink[0]{}%
	\providecommand \url  [0]{\begingroup\@sanitize@url \@url }%
	\providecommand \@url [1]{\endgroup\@href {#1}{\urlprefix }}%
	\providecommand \urlprefix  [0]{URL }%
	\providecommand \Eprint [0]{\href }%
	\providecommand \doibase [0]{https://doi.org/}%
	\providecommand \selectlanguage [0]{\@gobble}%
	\providecommand \bibinfo  [0]{\@secondoftwo}%
	\providecommand \bibfield  [0]{\@secondoftwo}%
	\providecommand \translation [1]{[#1]}%
	\providecommand \BibitemOpen [0]{}%
	\providecommand \bibitemStop [0]{}%
	\providecommand \bibitemNoStop [0]{.\EOS\space}%
	\providecommand \EOS [0]{\spacefactor3000\relax}%
	\providecommand \BibitemShut  [1]{\csname bibitem#1\endcsname}%
	\let\auto@bib@innerbib\@empty
	\bibitem [{\citenamefont {Brunton}\ and\ \citenamefont
		{Kutz}(2019)}]{Brunton_Kutz_2019}%
	\BibitemOpen
	\bibfield  {author} {\bibinfo {author} {\bibfnamefont {S.~L.}\ \bibnamefont
			{Brunton}}\ and\ \bibinfo {author} {\bibfnamefont {J.~N.}\ \bibnamefont
			{Kutz}},\ }\href@noop {} {\emph {\bibinfo {title} {Data-Driven Science and
				Engineering: Machine Learning, Dynamical Systems, and Control}}}\ (\bibinfo
	{publisher} {Cambridge University Press},\ \bibinfo {year}
	{2019})\BibitemShut {NoStop}%
	\bibitem [{\citenamefont {Tang}\ \emph {et~al.}(2020)\citenamefont {Tang},
		\citenamefont {Kurths}, \citenamefont {Lin}, \citenamefont {Ott},\ and\
		\citenamefont {Kocarev}}]{machine_learning_and_chaos}%
	\BibitemOpen
	\bibfield  {author} {\bibinfo {author} {\bibfnamefont {Y.}~\bibnamefont
			{Tang}}, \bibinfo {author} {\bibfnamefont {J.}~\bibnamefont {Kurths}},
		\bibinfo {author} {\bibfnamefont {W.}~\bibnamefont {Lin}}, \bibinfo {author}
		{\bibfnamefont {E.}~\bibnamefont {Ott}},\ and\ \bibinfo {author}
		{\bibfnamefont {L.}~\bibnamefont {Kocarev}},\ }\bibfield  {title} {\bibinfo
		{title} {{Introduction to Focus Issue: When machine learning meets complex
				systems: Networks, chaos, and nonlinear dynamics}},\ }\href@noop {}
	{\bibfield  {journal} {\bibinfo  {journal} {Chaos: An Interdisciplinary
				Journal of Nonlinear Science}\ }\textbf {\bibinfo {volume} {30}},\ \bibinfo
		{pages} {063151} (\bibinfo {year} {2020})}\BibitemShut {NoStop}%
	\bibitem [{\citenamefont {Jaeger}(2001)}]{Jaeger2001}%
	\BibitemOpen
	\bibfield  {author} {\bibinfo {author} {\bibfnamefont {H.}~\bibnamefont
			{Jaeger}},\ }\bibfield  {title} {\bibinfo {title} {The''echo state''approach
			to analysing and training recurrent neural networks}\ }(\bibinfo {year}
	{2001})\BibitemShut {NoStop}%
	\bibitem [{\citenamefont {Maass}\ \emph {et~al.}(2002)\citenamefont {Maass},
		\citenamefont {Natschläger},\ and\ \citenamefont {Markram}}]{Maass_2002}%
	\BibitemOpen
	\bibfield  {author} {\bibinfo {author} {\bibfnamefont {W.}~\bibnamefont
			{Maass}}, \bibinfo {author} {\bibfnamefont {T.}~\bibnamefont
			{Natschläger}},\ and\ \bibinfo {author} {\bibfnamefont {H.}~\bibnamefont
			{Markram}},\ }\bibfield  {title} {\bibinfo {title} {Real-time computing
			without stable states: A new framework for neural computation based on
			perturbations},\ }\href {https://doi.org/10.1162/089976602760407955}
	{\bibfield  {journal} {\bibinfo  {journal} {Neural Computation}\ }\textbf
		{\bibinfo {volume} {14}},\ \bibinfo {pages} {2531} (\bibinfo {year}
		{2002})}\BibitemShut {NoStop}%
	\bibitem [{\citenamefont {Prokhorov}(2005)}]{Prokhorov_2005}%
	\BibitemOpen
	\bibfield  {author} {\bibinfo {author} {\bibfnamefont {D.}~\bibnamefont
			{Prokhorov}},\ }\bibfield  {title} {\bibinfo {title} {Echo state networks:
			appeal and challenges},\ }in\ \href
	{https://doi.org/10.1109/IJCNN.2005.1556091} {\emph {\bibinfo {booktitle}
			{Proceedings. 2005 IEEE International Joint Conference on Neural Networks,
				2005.}}},\ Vol.~\bibinfo {volume} {3}\ (\bibinfo {year} {2005})\ pp.\
	\bibinfo {pages} {1463--1466}\BibitemShut {NoStop}%
	\bibitem [{\citenamefont {Lu}\ \emph {et~al.}(2017)\citenamefont {Lu},
		\citenamefont {Pathak}, \citenamefont {Hunt}, \citenamefont {Girvan},
		\citenamefont {Brockett},\ and\ \citenamefont {Ott}}]{Lu_2017}%
	\BibitemOpen
	\bibfield  {author} {\bibinfo {author} {\bibfnamefont {Z.}~\bibnamefont
			{Lu}}, \bibinfo {author} {\bibfnamefont {J.}~\bibnamefont {Pathak}}, \bibinfo
		{author} {\bibfnamefont {B.}~\bibnamefont {Hunt}}, \bibinfo {author}
		{\bibfnamefont {M.}~\bibnamefont {Girvan}}, \bibinfo {author} {\bibfnamefont
			{R.}~\bibnamefont {Brockett}},\ and\ \bibinfo {author} {\bibfnamefont
			{E.}~\bibnamefont {Ott}},\ }\bibfield  {title} {\bibinfo {title} {{Reservoir
				observers: Model-free inference of unmeasured variables in chaotic
				systems}},\ }\href {https://doi.org/10.1063/1.4979665} {\bibfield  {journal}
		{\bibinfo  {journal} {Chaos: An Interdisciplinary Journal of Nonlinear
				Science}\ }\textbf {\bibinfo {volume} {27}},\ \bibinfo {pages} {041102}
		(\bibinfo {year} {2017})}\BibitemShut {NoStop}%
	\bibitem [{\citenamefont {Pathak}\ \emph {et~al.}(2017)\citenamefont {Pathak},
		\citenamefont {Lu}, \citenamefont {Hunt}, \citenamefont {Girvan},\ and\
		\citenamefont {Ott}}]{Pathak_2017}%
	\BibitemOpen
	\bibfield  {author} {\bibinfo {author} {\bibfnamefont {J.}~\bibnamefont
			{Pathak}}, \bibinfo {author} {\bibfnamefont {Z.}~\bibnamefont {Lu}}, \bibinfo
		{author} {\bibfnamefont {B.}~\bibnamefont {Hunt}}, \bibinfo {author}
		{\bibfnamefont {M.}~\bibnamefont {Girvan}},\ and\ \bibinfo {author}
		{\bibfnamefont {E.}~\bibnamefont {Ott}},\ }\bibfield  {title} {\bibinfo
		{title} {Using machine learning to replicate chaotic attractors and calculate
			lyapunov exponents from data},\ }\href {https://doi.org/10.1063/1.5010300}
	{\bibfield  {journal} {\bibinfo  {journal} {Chaos: An Interdisciplinary
				Journal of Nonlinear Science}\ }\textbf {\bibinfo {volume} {27}} (\bibinfo
		{year} {2017})}\BibitemShut {NoStop}%
	\bibitem [{\citenamefont {Pathak}\ \emph
		{et~al.}(2018{\natexlab{a}})\citenamefont {Pathak}, \citenamefont {Hunt},
		\citenamefont {Girvan}, \citenamefont {Lu},\ and\ \citenamefont
		{Ott}}]{Pathak_2018}%
	\BibitemOpen
	\bibfield  {author} {\bibinfo {author} {\bibfnamefont {J.}~\bibnamefont
			{Pathak}}, \bibinfo {author} {\bibfnamefont {B.}~\bibnamefont {Hunt}},
		\bibinfo {author} {\bibfnamefont {M.}~\bibnamefont {Girvan}}, \bibinfo
		{author} {\bibfnamefont {Z.}~\bibnamefont {Lu}},\ and\ \bibinfo {author}
		{\bibfnamefont {E.}~\bibnamefont {Ott}},\ }\bibfield  {title} {\bibinfo
		{title} {Model-free prediction of large spatiotemporally chaotic systems from
			data: A reservoir computing approach},\ }\href
	{https://doi.org/10.1103/PhysRevLett.120.024102} {\bibfield  {journal}
		{\bibinfo  {journal} {Phys. Rev. Lett.}\ }\textbf {\bibinfo {volume} {120}},\
		\bibinfo {pages} {024102} (\bibinfo {year} {2018}{\natexlab{a}})}\BibitemShut
	{NoStop}%
	\bibitem [{\citenamefont {Lu}\ \emph {et~al.}(2018)\citenamefont {Lu},
		\citenamefont {Hunt},\ and\ \citenamefont {Ott}}]{Lu_2018}%
	\BibitemOpen
	\bibfield  {author} {\bibinfo {author} {\bibfnamefont {Z.}~\bibnamefont
			{Lu}}, \bibinfo {author} {\bibfnamefont {B.~R.}\ \bibnamefont {Hunt}},\ and\
		\bibinfo {author} {\bibfnamefont {E.}~\bibnamefont {Ott}},\ }\bibfield
	{title} {\bibinfo {title} {{Attractor reconstruction by machine learning}},\
	}\href {https://doi.org/10.1063/1.5039508} {\bibfield  {journal} {\bibinfo
			{journal} {Chaos: An Interdisciplinary Journal of Nonlinear Science}\
		}\textbf {\bibinfo {volume} {28}},\ \bibinfo {pages} {061104} (\bibinfo
		{year} {2018})}\BibitemShut {NoStop}%
	\bibitem [{\citenamefont {Zimmermann}\ and\ \citenamefont
		{Parlitz}(2018)}]{Parlitz_2018}%
	\BibitemOpen
	\bibfield  {author} {\bibinfo {author} {\bibfnamefont {R.~S.}\ \bibnamefont
			{Zimmermann}}\ and\ \bibinfo {author} {\bibfnamefont {U.}~\bibnamefont
			{Parlitz}},\ }\bibfield  {title} {\bibinfo {title} {{Observing
				spatio-temporal dynamics of excitable media using reservoir computing}},\
	}\href {https://doi.org/10.1063/1.5022276} {\bibfield  {journal} {\bibinfo
			{journal} {Chaos: An Interdisciplinary Journal of Nonlinear Science}\
		}\textbf {\bibinfo {volume} {28}},\ \bibinfo {pages} {043118} (\bibinfo
		{year} {2018})}\BibitemShut {NoStop}%
	\bibitem [{\citenamefont {Kong}\ \emph {et~al.}(2024)\citenamefont {Kong},
		\citenamefont {Brewer},\ and\ \citenamefont {Lai}}]{Kong_2024}%
	\BibitemOpen
	\bibfield  {author} {\bibinfo {author} {\bibfnamefont {L.-W.}\ \bibnamefont
			{Kong}}, \bibinfo {author} {\bibfnamefont {G.}~\bibnamefont {Brewer}},\ and\
		\bibinfo {author} {\bibfnamefont {Y.-C.}\ \bibnamefont {Lai}},\ }\bibfield
	{title} {\bibinfo {title} {Reservoir-computing based associative memory and
			itinerancy for complex dynamical attractors},\ }\href
	{https://doi.org/10.1038/s41467-024-49190-4} {\bibfield  {journal} {\bibinfo
			{journal} {Nature Communications}\ }\textbf {\bibinfo {volume} {15}}
		(\bibinfo {year} {2024})}\BibitemShut {NoStop}%
	\bibitem [{\citenamefont {Li}\ \emph {et~al.}(2024)\citenamefont {Li},
		\citenamefont {Zhu}, \citenamefont {Zhao}, \citenamefont {Duan},
		\citenamefont {Zhao}, \citenamefont {Zhang}, \citenamefont {Ma},
		\citenamefont {Sun},\ and\ \citenamefont {Lin}}]{Li_2024}%
	\BibitemOpen
	\bibfield  {author} {\bibinfo {author} {\bibfnamefont {X.}~\bibnamefont
			{Li}}, \bibinfo {author} {\bibfnamefont {Q.}~\bibnamefont {Zhu}}, \bibinfo
		{author} {\bibfnamefont {C.}~\bibnamefont {Zhao}}, \bibinfo {author}
		{\bibfnamefont {X.}~\bibnamefont {Duan}}, \bibinfo {author} {\bibfnamefont
			{B.}~\bibnamefont {Zhao}}, \bibinfo {author} {\bibfnamefont {X.}~\bibnamefont
			{Zhang}}, \bibinfo {author} {\bibfnamefont {H.-F.}\ \bibnamefont {Ma}},
		\bibinfo {author} {\bibfnamefont {J.}~\bibnamefont {Sun}},\ and\ \bibinfo
		{author} {\bibfnamefont {W.}~\bibnamefont {Lin}},\ }\bibfield  {title}
	{\bibinfo {title} {Higher-order granger reservoir computing: Simultaneously
			achieving scalable complex structures inference and accurate dynamics
			prediction},\ }\href {https://doi.org/10.1038/s41467-024-46852-1} {\bibfield
		{journal} {\bibinfo  {journal} {Nature Communications}\ }\textbf {\bibinfo
			{volume} {15}},\ \bibinfo {pages} {2506} (\bibinfo {year}
		{2024})}\BibitemShut {NoStop}%
	\bibitem [{\citenamefont {Fujii}\ and\ \citenamefont
		{Nakajima}(2017)}]{Fujii_2017}%
	\BibitemOpen
	\bibfield  {author} {\bibinfo {author} {\bibfnamefont {K.}~\bibnamefont
			{Fujii}}\ and\ \bibinfo {author} {\bibfnamefont {K.}~\bibnamefont
			{Nakajima}},\ }\bibfield  {title} {\bibinfo {title} {Harnessing
			disordered-ensemble quantum dynamics for machine learning},\ }\bibfield
	{journal} {\bibinfo  {journal} {Physical Review Applied}\ }\textbf {\bibinfo
		{volume} {8}},\ \href {https://doi.org/10.1103/physrevapplied.8.024030}
	{10.1103/physrevapplied.8.024030} (\bibinfo {year} {2017})\BibitemShut
	{NoStop}%
	\bibitem [{\citenamefont {Fujii}\ and\ \citenamefont
		{Nakajima}(2020)}]{Fujii_2020}%
	\BibitemOpen
	\bibfield  {author} {\bibinfo {author} {\bibfnamefont {K.}~\bibnamefont
			{Fujii}}\ and\ \bibinfo {author} {\bibfnamefont {K.}~\bibnamefont
			{Nakajima}},\ }\href@noop {} {\bibinfo {title} {Quantum reservoir computing:
			a reservoir approach toward quantum machine learning on near-term quantum
			devices}} (\bibinfo {year} {2020}),\ \Eprint
	{https://arxiv.org/abs/2011.04890} {arXiv:2011.04890 [quant-ph]} \BibitemShut
	{NoStop}%
	\bibitem [{\citenamefont {Nakajima}\ \emph {et~al.}(2019)\citenamefont
		{Nakajima}, \citenamefont {Fujii}, \citenamefont {Negoro}, \citenamefont
		{Mitarai},\ and\ \citenamefont {Kitagawa}}]{Nakajima_2019}%
	\BibitemOpen
	\bibfield  {author} {\bibinfo {author} {\bibfnamefont {K.}~\bibnamefont
			{Nakajima}}, \bibinfo {author} {\bibfnamefont {K.}~\bibnamefont {Fujii}},
		\bibinfo {author} {\bibfnamefont {M.}~\bibnamefont {Negoro}}, \bibinfo
		{author} {\bibfnamefont {K.}~\bibnamefont {Mitarai}},\ and\ \bibinfo {author}
		{\bibfnamefont {M.}~\bibnamefont {Kitagawa}},\ }\bibfield  {title} {\bibinfo
		{title} {Boosting computational power through spatial multiplexing in quantum
			reservoir computing},\ }\bibfield  {journal} {\bibinfo  {journal} {Physical
			Review Applied}\ }\textbf {\bibinfo {volume} {11}},\ \href
	{https://doi.org/10.1103/physrevapplied.11.034021}
	{10.1103/physrevapplied.11.034021} (\bibinfo {year} {2019})\BibitemShut
	{NoStop}%
	\bibitem [{\citenamefont {Mart\'{\i}nez-Pe\~na}\ \emph
		{et~al.}(2021)\citenamefont {Mart\'{\i}nez-Pe\~na}, \citenamefont {Giorgi},
		\citenamefont {Nokkala}, \citenamefont {Soriano},\ and\ \citenamefont
		{Zambrini}}]{Martinez_Pena}%
	\BibitemOpen
	\bibfield  {author} {\bibinfo {author} {\bibfnamefont {R.}~\bibnamefont
			{Mart\'{\i}nez-Pe\~na}}, \bibinfo {author} {\bibfnamefont {G.~L.}\
			\bibnamefont {Giorgi}}, \bibinfo {author} {\bibfnamefont {J.}~\bibnamefont
			{Nokkala}}, \bibinfo {author} {\bibfnamefont {M.~C.}\ \bibnamefont
			{Soriano}},\ and\ \bibinfo {author} {\bibfnamefont {R.}~\bibnamefont
			{Zambrini}},\ }\bibfield  {title} {\bibinfo {title} {Dynamical phase
			transitions in quantum reservoir computing},\ }\href
	{https://doi.org/10.1103/PhysRevLett.127.100502} {\bibfield  {journal}
		{\bibinfo  {journal} {Phys. Rev. Lett.}\ }\textbf {\bibinfo {volume} {127}},\
		\bibinfo {pages} {100502} (\bibinfo {year} {2021})}\BibitemShut {NoStop}%
	\bibitem [{\citenamefont {Chen}\ and\ \citenamefont
		{Nurdin}(2019)}]{Chen_2019}%
	\BibitemOpen
	\bibfield  {author} {\bibinfo {author} {\bibfnamefont {J.}~\bibnamefont
			{Chen}}\ and\ \bibinfo {author} {\bibfnamefont {H.~I.}\ \bibnamefont
			{Nurdin}},\ }\bibfield  {title} {\bibinfo {title} {Learning nonlinear
			input–output maps with dissipative quantum systems},\ }\bibfield  {journal}
	{\bibinfo  {journal} {Quantum Information Processing}\ }\textbf {\bibinfo
		{volume} {18}},\ \href {https://doi.org/10.1007/s11128-019-2311-9}
	{10.1007/s11128-019-2311-9} (\bibinfo {year} {2019})\BibitemShut {NoStop}%
	\bibitem [{\citenamefont {Martínez-Peña}\ \emph {et~al.}(2020)\citenamefont
		{Martínez-Peña}, \citenamefont {Nokkala}, \citenamefont {Giorgi},
		\citenamefont {Zambrini},\ and\ \citenamefont
		{Soriano}}]{Mart_nez_Pe_a_2020}%
	\BibitemOpen
	\bibfield  {author} {\bibinfo {author} {\bibfnamefont {R.}~\bibnamefont
			{Martínez-Peña}}, \bibinfo {author} {\bibfnamefont {J.}~\bibnamefont
			{Nokkala}}, \bibinfo {author} {\bibfnamefont {G.~L.}\ \bibnamefont {Giorgi}},
		\bibinfo {author} {\bibfnamefont {R.}~\bibnamefont {Zambrini}},\ and\
		\bibinfo {author} {\bibfnamefont {M.~C.}\ \bibnamefont {Soriano}},\
	}\bibfield  {title} {\bibinfo {title} {Information processing capacity of
			spin-based quantum reservoir computing systems},\ }\href
	{https://doi.org/10.1007/s12559-020-09772-y} {\bibfield  {journal} {\bibinfo
			{journal} {Cognitive Computation}\ }\textbf {\bibinfo {volume} {15}},\
		\bibinfo {pages} {1440–1451} (\bibinfo {year} {2020})}\BibitemShut
	{NoStop}%
	\bibitem [{\citenamefont {Tran}\ and\ \citenamefont
		{Nakajima}(2020)}]{tran2020}%
	\BibitemOpen
	\bibfield  {author} {\bibinfo {author} {\bibfnamefont {Q.~H.}\ \bibnamefont
			{Tran}}\ and\ \bibinfo {author} {\bibfnamefont {K.}~\bibnamefont
			{Nakajima}},\ }\href {https://arxiv.org/abs/2006.08999} {\bibinfo {title}
		{Higher-order quantum reservoir computing}} (\bibinfo {year} {2020}),\
	\Eprint {https://arxiv.org/abs/2006.08999} {arXiv:2006.08999 [quant-ph]}
	\BibitemShut {NoStop}%
	\bibitem [{\citenamefont {Domingo}\ \emph {et~al.}(2023)\citenamefont
		{Domingo}, \citenamefont {Grande}, \citenamefont {Carlo}, \citenamefont
		{Borondo},\ and\ \citenamefont {Borondo}}]{QRC_gate_computing}%
	\BibitemOpen
	\bibfield  {author} {\bibinfo {author} {\bibfnamefont {L.}~\bibnamefont
			{Domingo}}, \bibinfo {author} {\bibfnamefont {M.}~\bibnamefont {Grande}},
		\bibinfo {author} {\bibfnamefont {G.}~\bibnamefont {Carlo}}, \bibinfo
		{author} {\bibfnamefont {F.}~\bibnamefont {Borondo}},\ and\ \bibinfo {author}
		{\bibfnamefont {J.}~\bibnamefont {Borondo}},\ }\bibfield  {title} {\bibinfo
		{title} {{Optimal quantum reservoir computing for market forecasting: An
				application to fight food price crises}},\ }\href@noop {} {\  (\bibinfo
		{year} {2023})},\ \Eprint {https://arxiv.org/abs/2401.03347}
	{arXiv:2401.03347 [quant-ph]} \BibitemShut {NoStop}%
	\bibitem [{\citenamefont {Ghosh}\ \emph
		{et~al.}(2019{\natexlab{a}})\citenamefont {Ghosh}, \citenamefont {Opala},
		\citenamefont {Matuszewski}, \citenamefont {Paterek},\ and\ \citenamefont
		{Liew}}]{Ghosh_2019}%
	\BibitemOpen
	\bibfield  {author} {\bibinfo {author} {\bibfnamefont {S.}~\bibnamefont
			{Ghosh}}, \bibinfo {author} {\bibfnamefont {A.}~\bibnamefont {Opala}},
		\bibinfo {author} {\bibfnamefont {M.}~\bibnamefont {Matuszewski}}, \bibinfo
		{author} {\bibfnamefont {T.}~\bibnamefont {Paterek}},\ and\ \bibinfo {author}
		{\bibfnamefont {T.~C.~H.}\ \bibnamefont {Liew}},\ }\bibfield  {title}
	{\bibinfo {title} {Quantum reservoir processing},\ }\bibfield  {journal}
	{\bibinfo  {journal} {npj Quantum Information}\ }\textbf {\bibinfo {volume}
		{5}},\ \href {https://doi.org/10.1038/s41534-019-0149-8}
	{10.1038/s41534-019-0149-8} (\bibinfo {year}
	{2019}{\natexlab{a}})\BibitemShut {NoStop}%
	\bibitem [{\citenamefont {Ghosh}\ \emph
		{et~al.}(2019{\natexlab{b}})\citenamefont {Ghosh}, \citenamefont {Paterek},\
		and\ \citenamefont {Liew}}]{Ghosh2019QuantumNP}%
	\BibitemOpen
	\bibfield  {author} {\bibinfo {author} {\bibfnamefont {S.}~\bibnamefont
			{Ghosh}}, \bibinfo {author} {\bibfnamefont {T.}~\bibnamefont {Paterek}},\
		and\ \bibinfo {author} {\bibfnamefont {T.~C.~H.}\ \bibnamefont {Liew}},\
	}\bibfield  {title} {\bibinfo {title} {Quantum neuromorphic platform for
			quantum state preparation.},\ }\href
	{https://api.semanticscholar.org/CorpusID:210708242} {\bibfield  {journal}
		{\bibinfo  {journal} {Physical review letters}\ }\textbf {\bibinfo {volume}
			{123 26}},\ \bibinfo {pages} {260404} (\bibinfo {year}
		{2019}{\natexlab{b}})}\BibitemShut {NoStop}%
	\bibitem [{\citenamefont {Chen}\ \emph {et~al.}(2020)\citenamefont {Chen},
		\citenamefont {Nurdin},\ and\ \citenamefont {Yamamoto}}]{Chen_2020}%
	\BibitemOpen
	\bibfield  {author} {\bibinfo {author} {\bibfnamefont {J.}~\bibnamefont
			{Chen}}, \bibinfo {author} {\bibfnamefont {H.~I.}\ \bibnamefont {Nurdin}},\
		and\ \bibinfo {author} {\bibfnamefont {N.}~\bibnamefont {Yamamoto}},\
	}\bibfield  {title} {\bibinfo {title} {Temporal information processing on
			noisy quantum computers},\ }\bibfield  {journal} {\bibinfo  {journal}
		{Physical Review Applied}\ }\textbf {\bibinfo {volume} {14}},\ \href
	{https://doi.org/10.1103/physrevapplied.14.024065}
	{10.1103/physrevapplied.14.024065} (\bibinfo {year} {2020})\BibitemShut
	{NoStop}%
	\bibitem [{\citenamefont {Negoro}\ \emph {et~al.}(2018)\citenamefont {Negoro},
		\citenamefont {Mitarai}, \citenamefont {Fujii}, \citenamefont {Nakajima},\
		and\ \citenamefont {Kitagawa}}]{ML_nuclear_spin_ensemble}%
	\BibitemOpen
	\bibfield  {author} {\bibinfo {author} {\bibfnamefont {M.}~\bibnamefont
			{Negoro}}, \bibinfo {author} {\bibfnamefont {K.}~\bibnamefont {Mitarai}},
		\bibinfo {author} {\bibfnamefont {K.}~\bibnamefont {Fujii}}, \bibinfo
		{author} {\bibfnamefont {K.}~\bibnamefont {Nakajima}},\ and\ \bibinfo
		{author} {\bibfnamefont {M.}~\bibnamefont {Kitagawa}},\ }\href
	{https://arxiv.org/abs/1806.10910} {\bibinfo {title} {Machine learning with
			controllable quantum dynamics of a nuclear spin ensemble in a solid}}
	(\bibinfo {year} {2018}),\ \Eprint {https://arxiv.org/abs/1806.10910}
	{arXiv:1806.10910 [quant-ph]} \BibitemShut {NoStop}%
	\bibitem [{\citenamefont {Mujal}\ \emph {et~al.}(2021)\citenamefont {Mujal},
		\citenamefont {Mart{\'i}nez-Pe{\~n}a}, \citenamefont {Nokkala}, \citenamefont
		{Garc{\'i}a‐Beni}, \citenamefont {Giorgi}, \citenamefont {Soriano},\ and\
		\citenamefont {Zambrini}}]{Mujal2021OpportunitiesIQ}%
	\BibitemOpen
	\bibfield  {author} {\bibinfo {author} {\bibfnamefont {P.}~\bibnamefont
			{Mujal}}, \bibinfo {author} {\bibfnamefont {R.}~\bibnamefont
			{Mart{\'i}nez-Pe{\~n}a}}, \bibinfo {author} {\bibfnamefont {J.}~\bibnamefont
			{Nokkala}}, \bibinfo {author} {\bibfnamefont {J.}~\bibnamefont
			{Garc{\'i}a‐Beni}}, \bibinfo {author} {\bibfnamefont {G.~L.}\ \bibnamefont
			{Giorgi}}, \bibinfo {author} {\bibfnamefont {M.~C.}\ \bibnamefont
			{Soriano}},\ and\ \bibinfo {author} {\bibfnamefont {R.}~\bibnamefont
			{Zambrini}},\ }\bibfield  {title} {\bibinfo {title} {Opportunities in quantum
			reservoir computing and extreme learning machines},\ }\href
	{https://api.semanticscholar.org/CorpusID:232013896} {\bibfield  {journal}
		{\bibinfo  {journal} {Advanced Quantum Technologies}\ }\textbf {\bibinfo
			{volume} {4}} (\bibinfo {year} {2021})}\BibitemShut {NoStop}%
	\bibitem [{\citenamefont {Preskill}(2018)}]{Preskill2018}%
	\BibitemOpen
	\bibfield  {author} {\bibinfo {author} {\bibfnamefont {J.}~\bibnamefont
			{Preskill}},\ }\bibfield  {title} {\bibinfo {title} {Quantum {C}omputing in
			the {NISQ} era and beyond},\ }\href
	{https://doi.org/10.22331/q-2018-08-06-79} {\bibfield  {journal} {\bibinfo
			{journal} {{Quantum}}\ }\textbf {\bibinfo {volume} {2}},\ \bibinfo {pages}
		{79} (\bibinfo {year} {2018})}\BibitemShut {NoStop}%
	\bibitem [{\citenamefont {Bharti}\ \emph {et~al.}(2022)\citenamefont {Bharti},
		\citenamefont {Cervera-Lierta}, \citenamefont {Kyaw}, \citenamefont {Haug},
		\citenamefont {Alperin-Lea}, \citenamefont {Anand}, \citenamefont {Degroote},
		\citenamefont {Heimonen}, \citenamefont {Kottmann}, \citenamefont {Menke},
		\citenamefont {Mok}, \citenamefont {Sim}, \citenamefont {Kwek},\ and\
		\citenamefont {Aspuru-Guzik}}]{NISQ_era_algorithms}%
	\BibitemOpen
	\bibfield  {author} {\bibinfo {author} {\bibfnamefont {K.}~\bibnamefont
			{Bharti}}, \bibinfo {author} {\bibfnamefont {A.}~\bibnamefont
			{Cervera-Lierta}}, \bibinfo {author} {\bibfnamefont {T.~H.}\ \bibnamefont
			{Kyaw}}, \bibinfo {author} {\bibfnamefont {T.}~\bibnamefont {Haug}}, \bibinfo
		{author} {\bibfnamefont {S.}~\bibnamefont {Alperin-Lea}}, \bibinfo {author}
		{\bibfnamefont {A.}~\bibnamefont {Anand}}, \bibinfo {author} {\bibfnamefont
			{M.}~\bibnamefont {Degroote}}, \bibinfo {author} {\bibfnamefont
			{H.}~\bibnamefont {Heimonen}}, \bibinfo {author} {\bibfnamefont {J.~S.}\
			\bibnamefont {Kottmann}}, \bibinfo {author} {\bibfnamefont {T.}~\bibnamefont
			{Menke}}, \bibinfo {author} {\bibfnamefont {W.-K.}\ \bibnamefont {Mok}},
		\bibinfo {author} {\bibfnamefont {S.}~\bibnamefont {Sim}}, \bibinfo {author}
		{\bibfnamefont {L.-C.}\ \bibnamefont {Kwek}},\ and\ \bibinfo {author}
		{\bibfnamefont {A.}~\bibnamefont {Aspuru-Guzik}},\ }\bibfield  {title}
	{\bibinfo {title} {Noisy intermediate-scale quantum algorithms},\ }\bibfield
	{journal} {\bibinfo  {journal} {Reviews of Modern Physics}\ }\textbf
	{\bibinfo {volume} {94}},\ \href
	{https://doi.org/10.1103/revmodphys.94.015004} {10.1103/revmodphys.94.015004}
	(\bibinfo {year} {2022})\BibitemShut {NoStop}%
	\bibitem [{\citenamefont {Lau}\ \emph {et~al.}(2022)\citenamefont {Lau},
		\citenamefont {Lim}, \citenamefont {Shrotriya},\ and\ \citenamefont
		{Kwek}}]{Lau_2022}%
	\BibitemOpen
	\bibfield  {author} {\bibinfo {author} {\bibfnamefont {J.}~\bibnamefont
			{Lau}}, \bibinfo {author} {\bibfnamefont {K.}~\bibnamefont {Lim}}, \bibinfo
		{author} {\bibfnamefont {H.}~\bibnamefont {Shrotriya}},\ and\ \bibinfo
		{author} {\bibfnamefont {L.}~\bibnamefont {Kwek}},\ }\bibfield  {title}
	{\bibinfo {title} {Nisq computing: where are we and where do we go?},\ }\href
	{https://doi.org/10.1007/s43673-022-00058-z} {\bibfield  {journal} {\bibinfo
			{journal} {AAPPS Bulletin}\ }\textbf {\bibinfo {volume} {32}} (\bibinfo
		{year} {2022})}\BibitemShut {NoStop}%
	\bibitem [{\citenamefont {Ma}\ \emph {et~al.}(2023)\citenamefont {Ma},
		\citenamefont {Prosperino},\ and\ \citenamefont {Räth}}]{minimal_RC}%
	\BibitemOpen
	\bibfield  {author} {\bibinfo {author} {\bibfnamefont {H.}~\bibnamefont
			{Ma}}, \bibinfo {author} {\bibfnamefont {D.}~\bibnamefont {Prosperino}},\
		and\ \bibinfo {author} {\bibfnamefont {C.}~\bibnamefont {Räth}},\ }\bibfield
	{title} {\bibinfo {title} {A novel approach to minimal reservoir
			computing},\ }\href {https://doi.org/10.1038/s41598-023-39886-w} {\bibfield
		{journal} {\bibinfo  {journal} {Scientific Reports}\ }\textbf {\bibinfo
			{volume} {13}} (\bibinfo {year} {2023})}\BibitemShut {NoStop}%
	\bibitem [{\citenamefont {Akiba}\ \emph {et~al.}(2019)\citenamefont {Akiba},
		\citenamefont {Sano}, \citenamefont {Yanase}, \citenamefont {Ohta},\ and\
		\citenamefont {Koyama}}]{baysian_optimization_optuna}%
	\BibitemOpen
	\bibfield  {author} {\bibinfo {author} {\bibfnamefont {T.}~\bibnamefont
			{Akiba}}, \bibinfo {author} {\bibfnamefont {S.}~\bibnamefont {Sano}},
		\bibinfo {author} {\bibfnamefont {T.}~\bibnamefont {Yanase}}, \bibinfo
		{author} {\bibfnamefont {T.}~\bibnamefont {Ohta}},\ and\ \bibinfo {author}
		{\bibfnamefont {M.}~\bibnamefont {Koyama}},\ }\href@noop {} {\bibinfo {title}
		{Optuna: A next-generation hyperparameter optimization framework}} (\bibinfo
	{year} {2019}),\ \Eprint {https://arxiv.org/abs/1907.10902} {arXiv:1907.10902
		[cs.LG]} \BibitemShut {NoStop}%
	\bibitem [{\citenamefont {Duncan}\ and\ \citenamefont
		{Räth}(2023)}]{Hybrid_RC}%
	\BibitemOpen
	\bibfield  {author} {\bibinfo {author} {\bibfnamefont {D.}~\bibnamefont
			{Duncan}}\ and\ \bibinfo {author} {\bibfnamefont {C.}~\bibnamefont {Räth}},\
	}\bibfield  {title} {\bibinfo {title} {{Optimizing the combination of
				data-driven and model-based elements in hybrid reservoir computing}},\ }\href
	{https://doi.org/10.1063/5.0164013} {\bibfield  {journal} {\bibinfo
			{journal} {Chaos: An Interdisciplinary Journal of Nonlinear Science}\
		}\textbf {\bibinfo {volume} {33}},\ \bibinfo {pages} {103109} (\bibinfo
		{year} {2023})},\ \Eprint
	{https://arxiv.org/abs/https://pubs.aip.org/aip/cha/article-pdf/doi/10.1063/5.0164013/18159235/103109\_1\_5.0164013.pdf}
	{https://pubs.aip.org/aip/cha/article-pdf/doi/10.1063/5.0164013/18159235/103109\_1\_5.0164013.pdf}
	\BibitemShut {NoStop}%
	\bibitem [{\citenamefont {Giammarese}\ \emph {et~al.}(2024)\citenamefont
		{Giammarese}, \citenamefont {Rana}, \citenamefont {Bollt},\ and\
		\citenamefont {Malik}}]{giammarese2024}%
	\BibitemOpen
	\bibfield  {author} {\bibinfo {author} {\bibfnamefont {A.}~\bibnamefont
			{Giammarese}}, \bibinfo {author} {\bibfnamefont {K.}~\bibnamefont {Rana}},
		\bibinfo {author} {\bibfnamefont {E.~M.}\ \bibnamefont {Bollt}},\ and\
		\bibinfo {author} {\bibfnamefont {N.}~\bibnamefont {Malik}},\ }\href
	{https://arxiv.org/abs/2403.13836} {\bibinfo {title} {Tree-based learning for
			high-fidelity prediction of chaos}} (\bibinfo {year} {2024}),\ \Eprint
	{https://arxiv.org/abs/2403.13836} {arXiv:2403.13836 [cs.LG]} \BibitemShut
	{NoStop}%
	\bibitem [{\citenamefont {Haluszczynski}\ and\ \citenamefont
		{Räth}(2019)}]{RC_1}%
	\BibitemOpen
	\bibfield  {author} {\bibinfo {author} {\bibfnamefont {A.}~\bibnamefont
			{Haluszczynski}}\ and\ \bibinfo {author} {\bibfnamefont {C.}~\bibnamefont
			{Räth}},\ }\bibfield  {title} {\bibinfo {title} {{Good and bad predictions:
				Assessing and improving the replication of chaotic attractors by means of
				reservoir computing}},\ }\href {https://doi.org/10.1063/1.5118725} {\bibfield
		{journal} {\bibinfo  {journal} {Chaos: An Interdisciplinary Journal of
				Nonlinear Science}\ }\textbf {\bibinfo {volume} {29}},\ \bibinfo {pages}
		{103143} (\bibinfo {year} {2019})},\ \Eprint
	{https://arxiv.org/abs/https://pubs.aip.org/aip/cha/article-pdf/doi/10.1063/1.5118725/14626476/103143\_1\_online.pdf}
	{https://pubs.aip.org/aip/cha/article-pdf/doi/10.1063/1.5118725/14626476/103143\_1\_online.pdf}
	\BibitemShut {NoStop}%
	\bibitem [{\citenamefont {Haluszczynski}\ \emph {et~al.}(2020)\citenamefont
		{Haluszczynski}, \citenamefont {Aumeier}, \citenamefont {Herteux},\ and\
		\citenamefont {Räth}}]{RC_2}%
	\BibitemOpen
	\bibfield  {author} {\bibinfo {author} {\bibfnamefont {A.}~\bibnamefont
			{Haluszczynski}}, \bibinfo {author} {\bibfnamefont {J.}~\bibnamefont
			{Aumeier}}, \bibinfo {author} {\bibfnamefont {J.}~\bibnamefont {Herteux}},\
		and\ \bibinfo {author} {\bibfnamefont {C.}~\bibnamefont {Räth}},\ }\bibfield
	{title} {\bibinfo {title} {{Reducing network size and improving prediction
				stability of reservoir computing}},\ }\href
	{https://doi.org/10.1063/5.0006869} {\bibfield  {journal} {\bibinfo
			{journal} {Chaos: An Interdisciplinary Journal of Nonlinear Science}\
		}\textbf {\bibinfo {volume} {30}},\ \bibinfo {pages} {063136} (\bibinfo
		{year} {2020})},\ \Eprint
	{https://arxiv.org/abs/https://pubs.aip.org/aip/cha/article-pdf/doi/10.1063/5.0006869/14629823/063136\_1\_online.pdf}
	{https://pubs.aip.org/aip/cha/article-pdf/doi/10.1063/5.0006869/14629823/063136\_1\_online.pdf}
	\BibitemShut {NoStop}%
	\bibitem [{\citenamefont {Herteux}\ and\ \citenamefont {Räth}(2020)}]{RC_3}%
	\BibitemOpen
	\bibfield  {author} {\bibinfo {author} {\bibfnamefont {J.}~\bibnamefont
			{Herteux}}\ and\ \bibinfo {author} {\bibfnamefont {C.}~\bibnamefont
			{Räth}},\ }\bibfield  {title} {\bibinfo {title} {{Breaking symmetries of the
				reservoir equations in echo state networks}},\ }\href
	{https://doi.org/10.1063/5.0028993} {\bibfield  {journal} {\bibinfo
			{journal} {Chaos: An Interdisciplinary Journal of Nonlinear Science}\
		}\textbf {\bibinfo {volume} {30}},\ \bibinfo {pages} {123142} (\bibinfo
		{year} {2020})},\ \Eprint
	{https://arxiv.org/abs/https://pubs.aip.org/aip/cha/article-pdf/doi/10.1063/5.0028993/14110522/123142\_1\_online.pdf}
	{https://pubs.aip.org/aip/cha/article-pdf/doi/10.1063/5.0028993/14110522/123142\_1\_online.pdf}
	\BibitemShut {NoStop}%
	\bibitem [{\citenamefont {Pathak}\ \emph
		{et~al.}(2018{\natexlab{b}})\citenamefont {Pathak}, \citenamefont {Wikner},
		\citenamefont {Fussell}, \citenamefont {Chandra}, \citenamefont {Hunt},
		\citenamefont {Girvan},\ and\ \citenamefont {Ott}}]{forecast_horizon}%
	\BibitemOpen
	\bibfield  {author} {\bibinfo {author} {\bibfnamefont {J.}~\bibnamefont
			{Pathak}}, \bibinfo {author} {\bibfnamefont {A.}~\bibnamefont {Wikner}},
		\bibinfo {author} {\bibfnamefont {R.}~\bibnamefont {Fussell}}, \bibinfo
		{author} {\bibfnamefont {S.}~\bibnamefont {Chandra}}, \bibinfo {author}
		{\bibfnamefont {B.~R.}\ \bibnamefont {Hunt}}, \bibinfo {author}
		{\bibfnamefont {M.}~\bibnamefont {Girvan}},\ and\ \bibinfo {author}
		{\bibfnamefont {E.}~\bibnamefont {Ott}},\ }\bibfield  {title} {\bibinfo
		{title} {{Hybrid forecasting of chaotic processes: Using machine learning in
				conjunction with a knowledge-based model}},\ }\href
	{https://doi.org/10.1063/1.5028373} {\bibfield  {journal} {\bibinfo
			{journal} {Chaos: An Interdisciplinary Journal of Nonlinear Science}\
		}\textbf {\bibinfo {volume} {28}},\ \bibinfo {pages} {041101} (\bibinfo
		{year} {2018}{\natexlab{b}})},\ \Eprint
	{https://arxiv.org/abs/https://pubs.aip.org/aip/cha/article-pdf/doi/10.1063/1.5028373/10315097/041101\_1\_online.pdf}
	{https://pubs.aip.org/aip/cha/article-pdf/doi/10.1063/1.5028373/10315097/041101\_1\_online.pdf}
	\BibitemShut {NoStop}%
	\bibitem [{\citenamefont {Grassberger}\ and\ \citenamefont
		{Procaccia}(1983)}]{GRASSBERGER_correlation_dimension}%
	\BibitemOpen
	\bibfield  {author} {\bibinfo {author} {\bibfnamefont {P.}~\bibnamefont
			{Grassberger}}\ and\ \bibinfo {author} {\bibfnamefont {I.}~\bibnamefont
			{Procaccia}},\ }\bibfield  {title} {\bibinfo {title} {Measuring the
			strangeness of strange attractors},\ }\href
	{https://doi.org/https://doi.org/10.1016/0167-2789(83)90298-1} {\bibfield
		{journal} {\bibinfo  {journal} {Physica D: Nonlinear Phenomena}\ }\textbf
		{\bibinfo {volume} {9}},\ \bibinfo {pages} {189} (\bibinfo {year}
		{1983})}\BibitemShut {NoStop}%
	\bibitem [{\citenamefont {Grassberger}(1983)}]{Grassberger_Procaccia}%
	\BibitemOpen
	\bibfield  {author} {\bibinfo {author} {\bibfnamefont {P.}~\bibnamefont
			{Grassberger}},\ }\bibfield  {title} {\bibinfo {title} {Generalized
			dimensions of strange attractors},\ }\href
	{https://doi.org/https://doi.org/10.1016/0375-9601(83)90753-3} {\bibfield
		{journal} {\bibinfo  {journal} {Physics Letters A}\ }\textbf {\bibinfo
			{volume} {97}},\ \bibinfo {pages} {227} (\bibinfo {year} {1983})}\BibitemShut
	{NoStop}%
	\bibitem [{\citenamefont {Wolf}\ \emph {et~al.}(1985)\citenamefont {Wolf},
		\citenamefont {Swift}, \citenamefont {Swinney},\ and\ \citenamefont
		{Vastano}}]{Wolf_1985}%
	\BibitemOpen
	\bibfield  {author} {\bibinfo {author} {\bibfnamefont {A.}~\bibnamefont
			{Wolf}}, \bibinfo {author} {\bibfnamefont {J.}~\bibnamefont {Swift}},
		\bibinfo {author} {\bibfnamefont {H.~L.}\ \bibnamefont {Swinney}},\ and\
		\bibinfo {author} {\bibfnamefont {J.~A.}\ \bibnamefont {Vastano}},\
	}\bibfield  {title} {\bibinfo {title} {Determining lyapunov exponents from a
			time series},\ }\href {https://api.semanticscholar.org/CorpusID:14411384}
	{\bibfield  {journal} {\bibinfo  {journal} {Physica D: Nonlinear Phenomena}\
		}\textbf {\bibinfo {volume} {16}},\ \bibinfo {pages} {285} (\bibinfo {year}
		{1985})}\BibitemShut {NoStop}%
	\bibitem [{\citenamefont {Shaw}(1981)}]{Shaw}%
	\BibitemOpen
	\bibfield  {author} {\bibinfo {author} {\bibfnamefont {R.}~\bibnamefont
			{Shaw}},\ }\bibfield  {title} {\bibinfo {title} {Strange attractors, chaotic
			behavior, and information flow},\ }\href
	{https://doi.org/doi:10.1515/zna-1981-0115} {\bibfield  {journal} {\bibinfo
			{journal} {Zeitschrift für Naturforschung A}\ }\textbf {\bibinfo {volume}
			{36}},\ \bibinfo {pages} {80} (\bibinfo {year} {1981})}\BibitemShut {NoStop}%
	\bibitem [{\citenamefont {Rosenstein}\ \emph {et~al.}(1993)\citenamefont
		{Rosenstein}, \citenamefont {Collins},\ and\ \citenamefont {{De
				Luca}}}]{Rosenstein_algorithm}%
	\BibitemOpen
	\bibfield  {author} {\bibinfo {author} {\bibfnamefont {M.~T.}\ \bibnamefont
			{Rosenstein}}, \bibinfo {author} {\bibfnamefont {J.~J.}\ \bibnamefont
			{Collins}},\ and\ \bibinfo {author} {\bibfnamefont {C.~J.}\ \bibnamefont {{De
					Luca}}},\ }\bibfield  {title} {\bibinfo {title} {A practical method for
			calculating largest lyapunov exponents from small data sets},\ }\href
	{https://doi.org/https://doi.org/10.1016/0167-2789(93)90009-P} {\bibfield
		{journal} {\bibinfo  {journal} {Physica D: Nonlinear Phenomena}\ }\textbf
		{\bibinfo {volume} {65}},\ \bibinfo {pages} {117} (\bibinfo {year}
		{1993})}\BibitemShut {NoStop}%
\end{thebibliography}
%
\end{document}



\title{Supplemental information \\ Predicting three-dimensional chaotic systems with four qubit quantum systems}


\author{Joel Steinegger}
\email{J.Steinegger@t-online.de}
\affiliation{Ludwig-Maximilians-Universit\"at, Department of Physics, Schellingstra{\ss}e 4, 80799 Munich, Germany}
\affiliation{Deutsches Zentrum f{\"u}r Luft- und Raumfahrt (DLR), Institut f{\"u}r KI Sicherheit, Wilhelm-Runge-Stra{\ss}e 10, 89081 Ulm, Germany}

\author{Christoph R\"ath}
\email{christoph.raeth@dlr.de}
\affiliation{Deutsches Zentrum f{\"u}r Luft- und Raumfahrt (DLR), Institut f{\"u}r KI Sicherheit, Wilhelm-Runge-Stra{\ss}e 10, 89081 Ulm, Germany}

\date{\today}

\begin{center}
  \large
  \textbf{Supplemental information \\ Predicting three-dimensional chaotic systems with four qubit quantum systems} \\
  Joel Steinegger$^{1,2}$  and Christoph Räth$^{1, \dagger}$ \\[0.5em]
  \begin{tiny}
  $^{1}$ Deutsches Zentrum f{\"u}r Luft- und Raumfahrt (DLR), Institut f{\"u}r KI Sicherheit, Wilhelm-Runge-Stra{\ss}e 10, 89081 Ulm, Germany\\
  $^{2}$ Institut f{\"u}r Materialphysik im Weltraum, Deutsches Zentrum für Luft- und Raumfahrt (DLR), 51170 K{\"o}ln, Germany\\
  $^{\dagger}$Correspondence to: christoph.raeth@dlr.de \\
  January 25, 2025
  \end{tiny}
\end{center}

\section{\label{sec:non_trivial_hyperspace}Supplementary Note 1: Non-trivial Hyperparameterspace}
\begin{figure*}[h!]
    \begin{center}
    \includegraphics[width=\textwidth]{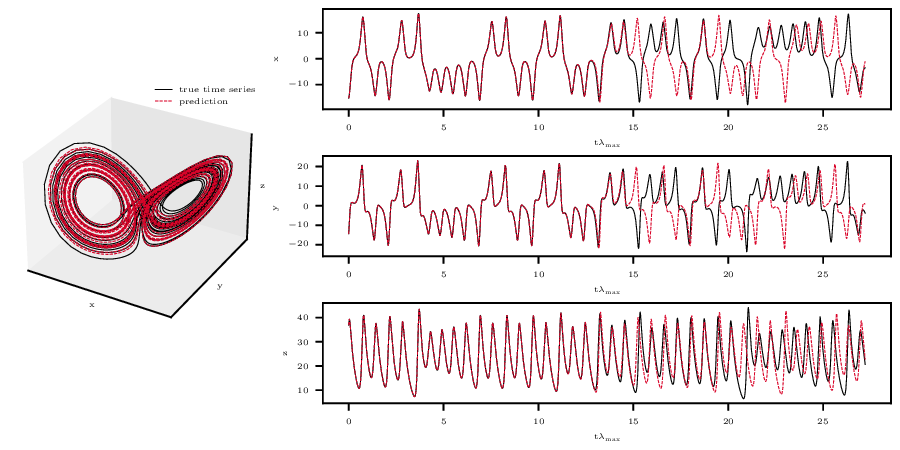}
    \end{center}
    \caption{\label{fig:lorenz_trajectory_best} Example of a trajectory of one of the 500 continued time series (TABLE \ref{table:hyperparameter_choices}) obtained for the Lorenz attractor.}
\end{figure*}

\begin{figure*}[h!]
  \centering
  \begin{minipage}[b]{0.48\textwidth}
    \centering
    \includegraphics[width=\textwidth]{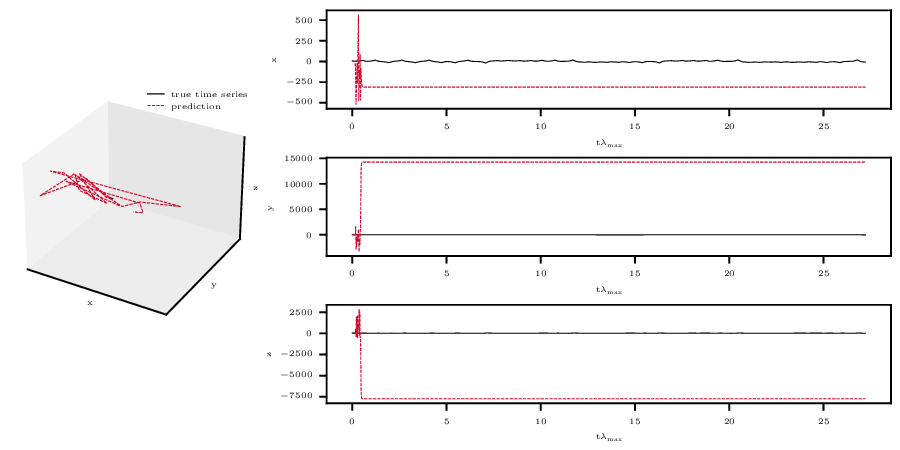}
    \caption{\label{fig:lorenz_trajectory_small_reg} Example of a trajectory of the prediction phase with a very small regression parameter.}
  \end{minipage}
  \hfill
  \begin{minipage}[b]{0.48\textwidth}
    \centering
    \includegraphics[width=\textwidth]{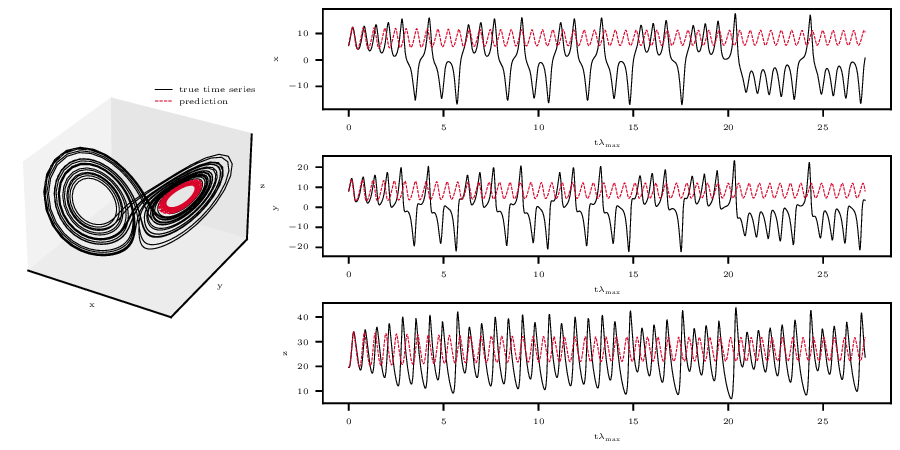}
    \caption{\label{fig:lorenz_trajectory_large_reg}Example of a trajectory of the prediction phase with a very large regression parameter.}
  \end{minipage}
\end{figure*}

\begin{figure*}[h!]
    \begin{center}
    \includegraphics[width=0.9\textwidth]{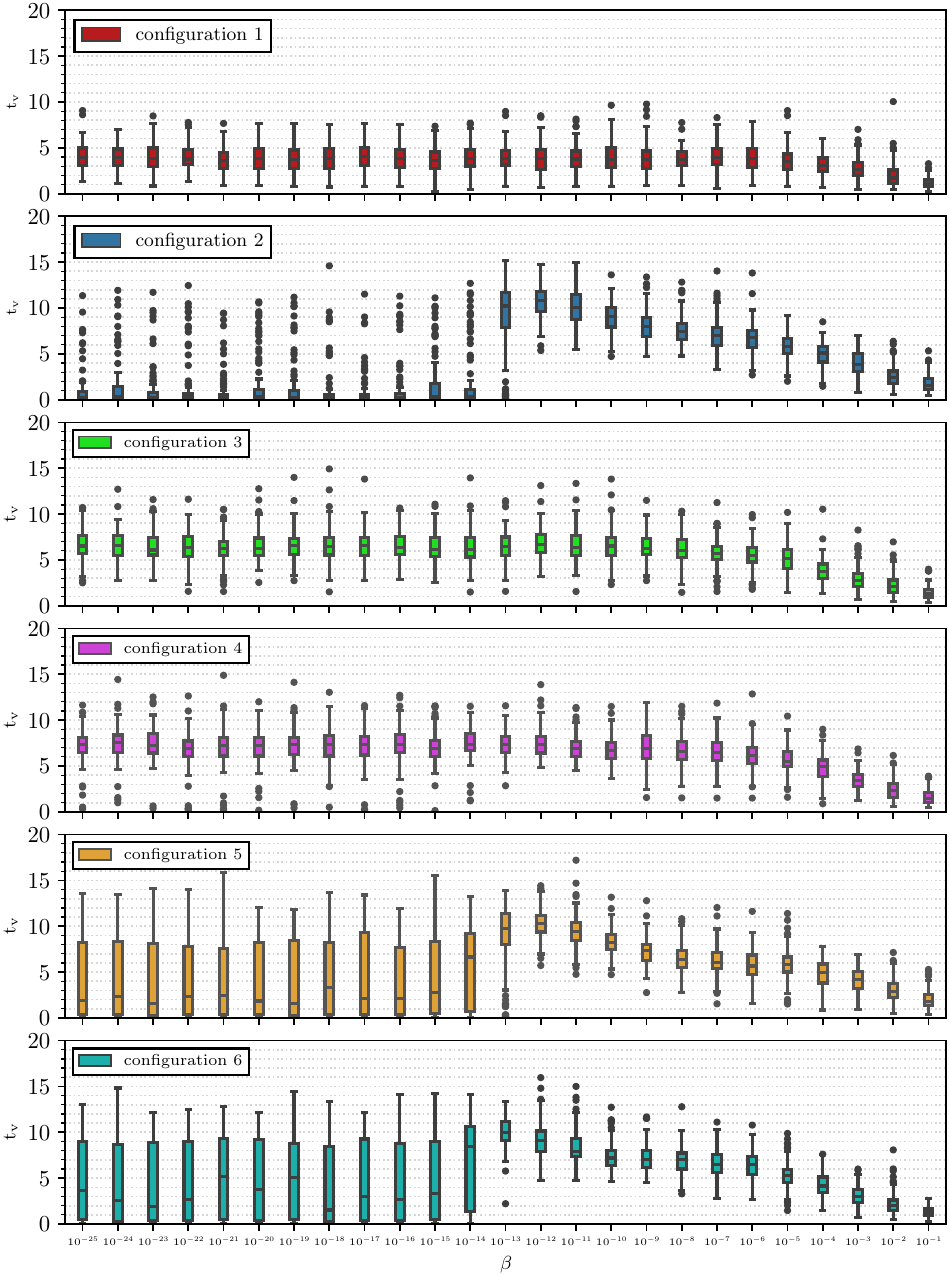}
    \end{center}
    \caption{\label{fig:forecast_regression_parameters}Boxplot of the forecast horizon in Lyapunov times of 100 forecasted trajectories of the Lorenz-63 system against the regression parameter ($\beta \in \{10^{-1}, 10^{-3}, 10^{-10}, 10^{-20} \}$) for six different hyperparamter configurations of $r, G, V$ and $[a, b]$ (see Table \ref{table:configurations_ridge_regression}).}
\end{figure*}

\begin{figure*}[h!]
    \begin{center}
    \includegraphics[width=0.9\textwidth]{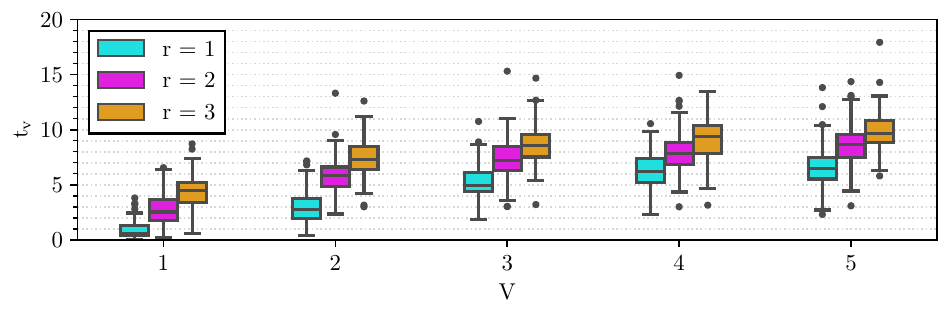}
    \end{center}
    \caption{\label{fig:reservoirs_sweep_forecast_horizon}Boxplot (100 forecasted trajectories) presenting the forecast horizon of the prediction of the Lorenz system with $[a,b]$=[0.2,0.8], $\beta=10^{-10}$, $G=1$ and varying $V$ and $r$.}
\end{figure*}

\begin{figure*}
    \begin{center}
    \includegraphics[width=0.9\textwidth]{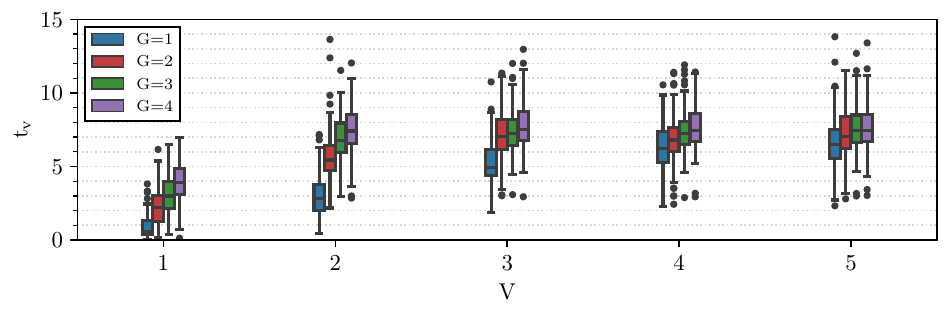}
    \end{center}
    \caption{\label{fig:fcts_sweep_forecast_horizon} Boxplot (100 forecasted trajectories) presenting the forecast horizon of the prediction of the Lorenz system with $[a,b]$=[0.2,0.8], $\beta=10^{-10}$, $r$=1 and varying $V$ and $G$.}
\end{figure*}

\begin{figure*}[h]
    \begin{center}
    \includegraphics[width=0.9\textwidth]{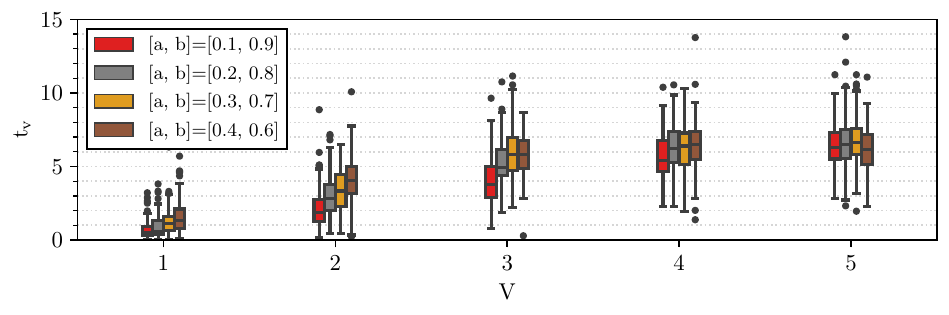}
    \end{center}
    \caption{\label{fig:interval_sweep_forecast_horizon} Boxplot (100 forecasted trajectories) presenting the forecast horizon of the prediction of the Lorenz system with $G$=1, $\beta=10^{-10}$, $r$=1 and varying $V$ and $[a,b]$.}
\end{figure*}

\begin{figure*}
    \begin{center}
    \includegraphics[width=0.9\textwidth]{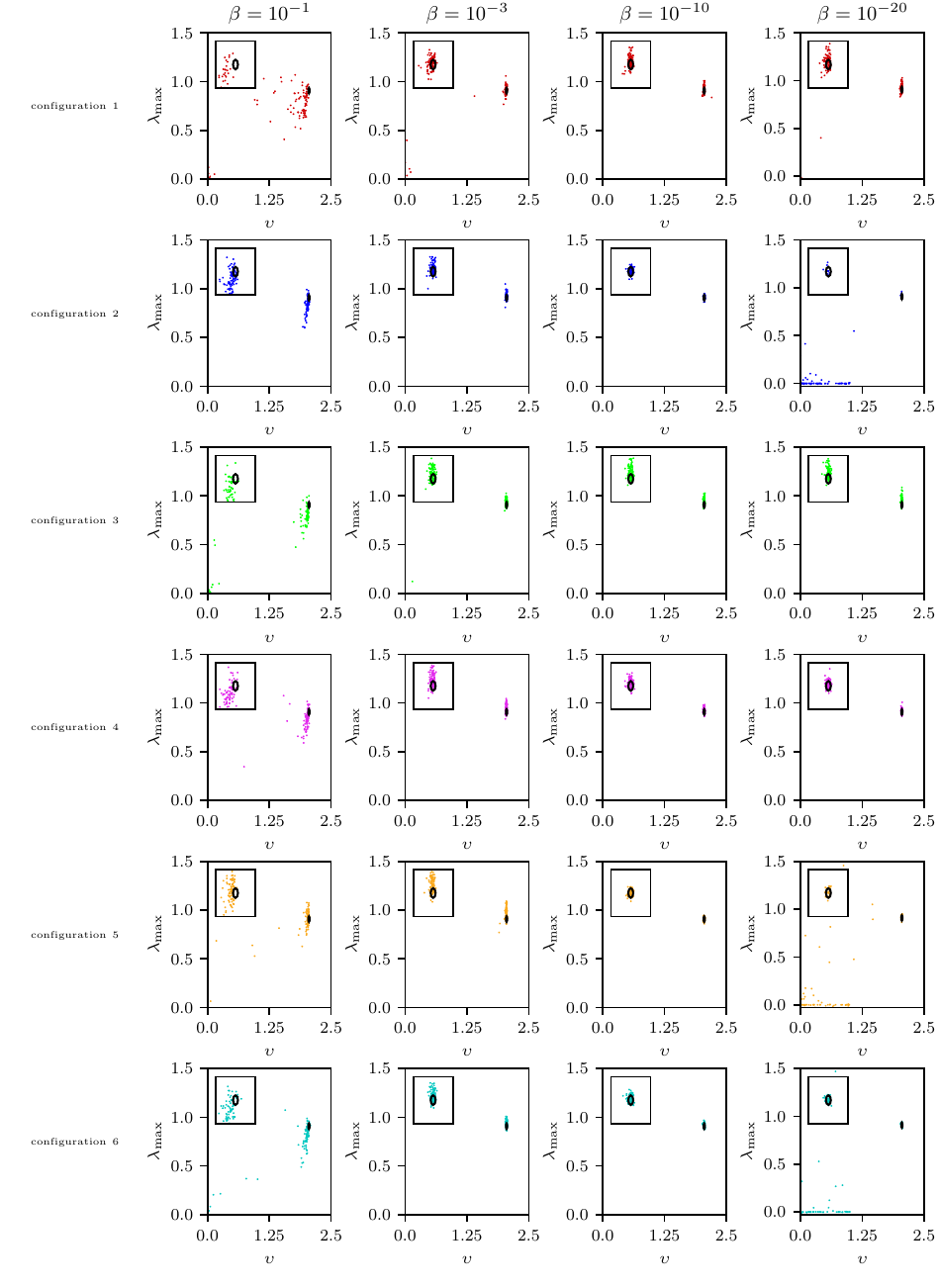}
    \end{center}
    \caption{\label{fig:climate_regression_parameters}Scatter plots of the predicted forecast horizons showing the predicted $\lambda_{\mathrm{max}}$ scattered against the predicted $\upsilon$ for each of the 100 realizations for $\beta \in \{ 10^{-1}, 10^{-3}, 10^{-10}, 10^{-20}\}$ and six different configurations of $r, V, G$ and $[a, b]$ (see Table \ref{table:configurations_ridge_regression}).}
\end{figure*}

\begin{figure*}
    \begin{center}
    \includegraphics[width=\textwidth]{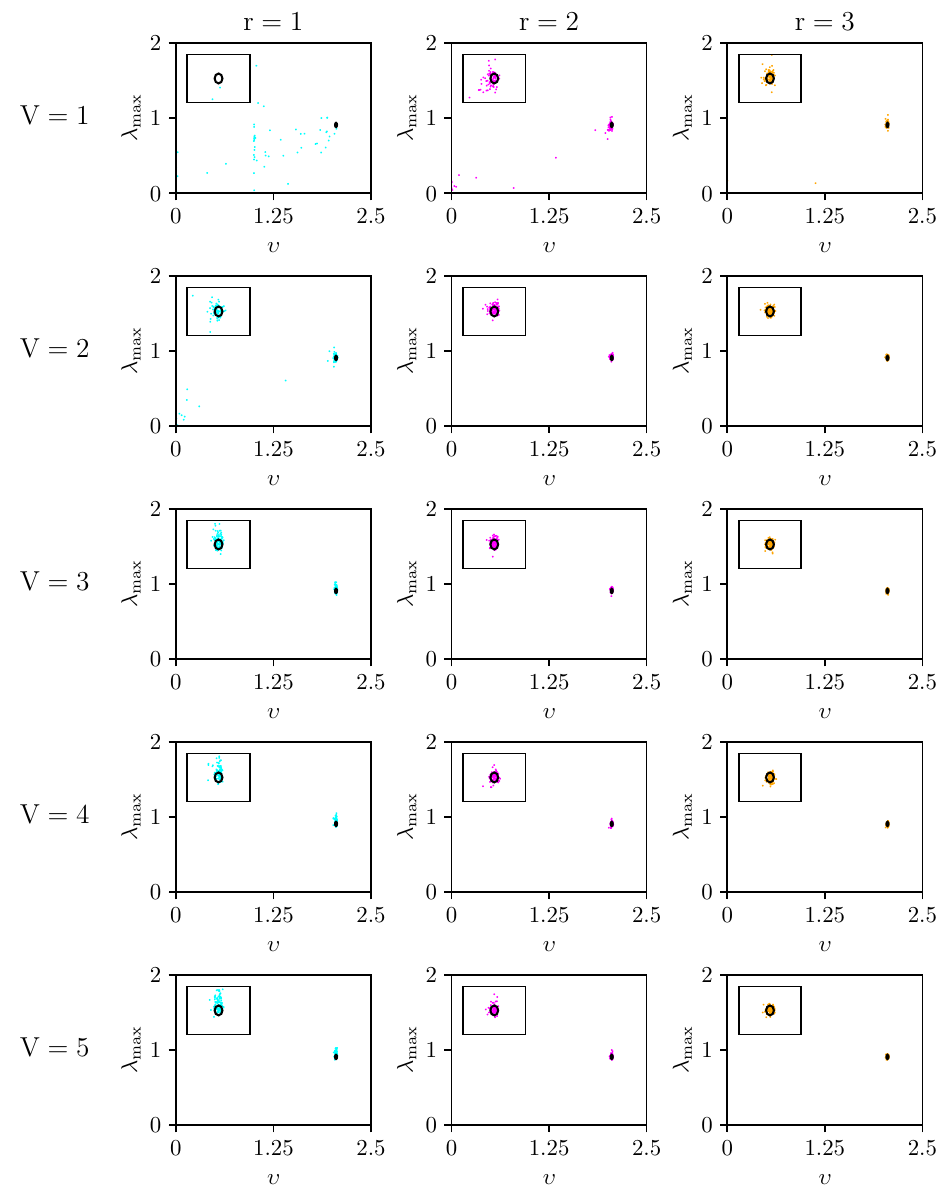}
    \end{center}
    \caption{\label{fig:reservoirs_sweep_climate}  Largest Lyapunov exponent scattered against correlation dimension 100 forecasted trajectories for each of the hyperparameter combinations varying $V$ and $r$ while holding [$a$,$b$]=[0.2,0.8], $\beta=10^{-10}$ and $G=1$ fixed.}
\end{figure*}

\begin{figure*}
    \begin{center}
    \includegraphics[width=\textwidth]{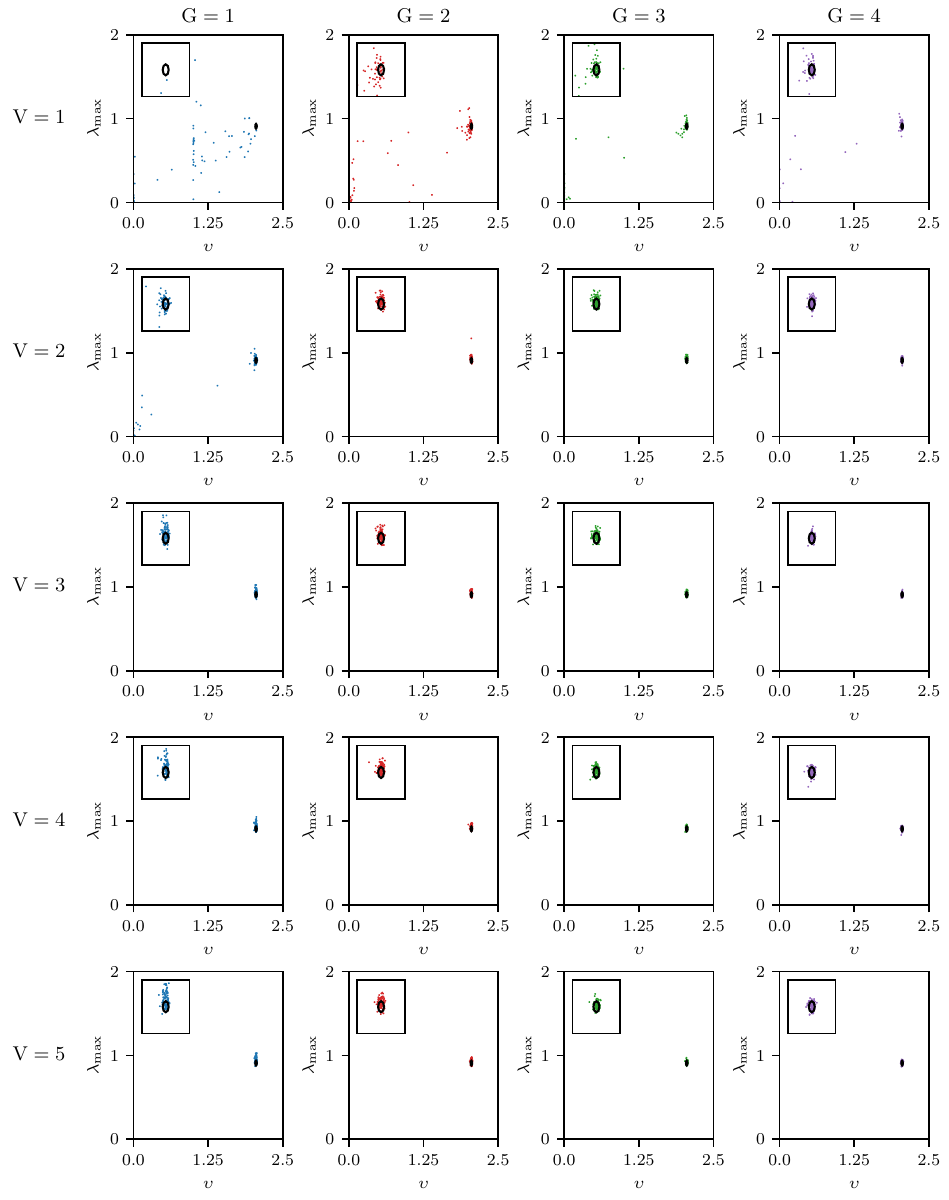}
    \end{center}
    \caption{\label{fig:fcts_sweep_climate} Largest Lyapunov exponent scattered against correlation dimension 100 forecasted trajectories for each of the hyperparameter combinations varying $V$ and $G$ while holding [$a$,$b$]=[0.2,0.8], $\beta=10^{-10}$, $r=1$ fixed.}
\end{figure*}

\begin{figure*}[h!]
    \begin{center}
    \includegraphics[width=\textwidth]{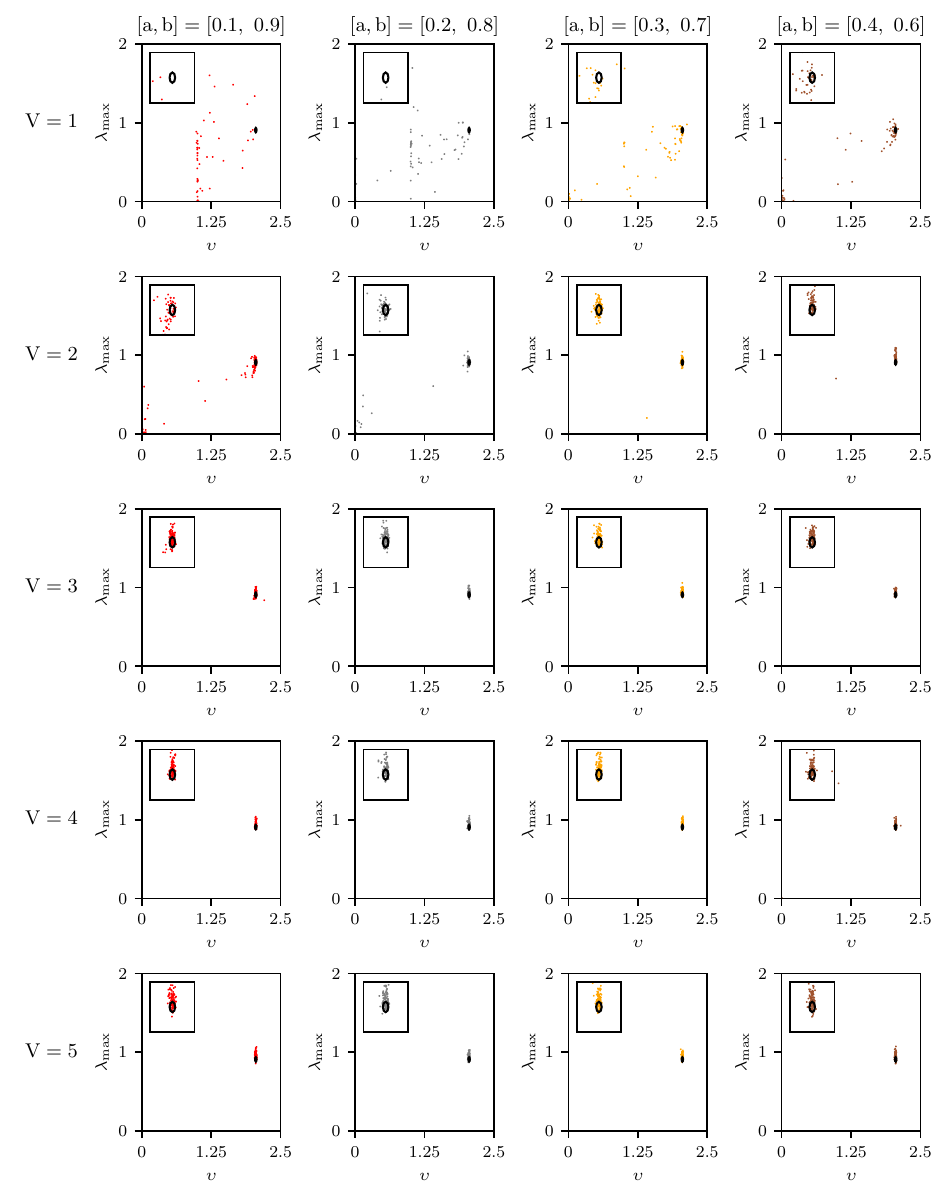}
    \end{center}
    \caption{\label{fig:intervals_sweep_climate} Largest Lyapunov exponent scattered against correlation dimension 100 forecasted trajectories for each of the hyperparameter combinations varying varying $V$ and [$a$,$b$] while $G=1$, $\beta=10^{-10}$ and $r=1$ fixed.}
\end{figure*}

In the following, we want to motivate the choice of free hyperparameters by showing the results of small and thoughtfully chosen parameter sweeps. They are obtained with $N_{\mathrm{stat}}$=100 for each hyperparameter combination. In the original QRC algorithm \cite{Fujii_2017} the Moore–Penrose-inverse is calculated to obtain the readout matrix. In this work, we use ridge regression instead of the pseudo inverse to counteract overfitting by adding a term that penalizes large matrix coefficients to the loss function. This is the common approach in "classical" RC. It introduces the regression parameter $\beta$ that is critical to optimize. The importance of the introduction and optimization of $\beta$ is analyzed by training the models for all $\beta \in \{10^{-25}, 10^{-24}, 10^{-23},... , 10^{-2},10^{-1}\}$ for 6 parameter configurations  (see Table \ref{table:configurations_ridge_regression}). The obtained forecast horizons are depicted in Fig. \ref{fig:forecast_regression_parameters} and the predicted climate is shown for $\beta \in \{10^{-1}, 10^{-3}, 10^{-10}, 10^{-20} \}$ in Fig. \ref{fig:climate_regression_parameters}. Two important observations can be made. Firstly, it can be observed that good short- and long-term prediction quality depends critically on the right choice of $\beta$. Secondly, it can be seen that the importance of the optimization of the regression parameter seems to grow with the dimension of the output vector $\mathrm{\mathbf{q}}(k)$. The first observation is further illustrated with the following example trajectories: Fig. \ref{fig:lorenz_trajectory_best} is an example of one of the prediction phases of the best performing hyperparameter combination (Table \ref{table:hyperparameter_choices}) of the Lorenz-63 attractor presented for 1500 prediction steps. Changing the regression parameter to a very small one ($\beta=10^{-24}$) frequently results in diverging trajectories, like in Fig. \ref{fig:lorenz_trajectory_small_reg}. A very large regression parameter ($\beta=10$) typically leads to periodic predictions like in Fig. \ref{fig:lorenz_trajectory_large_reg}. These extreme parameter choices result in models that we can characterize as not working. This illustrates that the regression parameter has to be carefully optimized to achieve good prediction performance. The increase of the reservoir output dimension and its effect on the short-term prediction quality are illustrated in Fig. \ref{fig:reservoirs_sweep_forecast_horizon} and Fig. \ref{fig:fcts_sweep_forecast_horizon} and its effect on the long-term climate is shown in Fig. \ref{fig:reservoirs_sweep_climate} and Fig. \ref{fig:fcts_sweep_climate}. The forecasted continuation of the trajectories matches the actual continuation of the time series very closely for longer times when the dimension of the output vector $\mathrm{\mathbf{q}}(k)$ is increased. The prediction of the climate patterns (especially the prediction of the largest Lyapunov exponent) increases with the output dimension of the output vector at step $k$. The remaining free hyperparameter is the scaling interval [$a$,$b$]. Its effects on the short- and long-term behavior are shown in Fig. \ref{fig:interval_sweep_forecast_horizon} and Fig. \ref{fig:intervals_sweep_climate}. The right choice of [$a$,$b$] seems to critically depend on the other hyperparameters. 
\begin{table}[h]
\centering
\resizebox{\linewidth}{!}{
\begin{tabular}{|l|c|c|c|c|c|}
    \hline
    configuration label & \hspace{0.5cm} $V$ \hspace{0.5cm} & \hspace{0.5cm} $r$ \hspace{0.5cm} & \hspace{0.5cm} $G$ \hspace{0.5cm} & \hspace{0.5cm} [$a$, $b$] \hspace{0.5cm} & \hspace{0.5cm} dim($\mathrm{\mathbf{q}}(k)$) \hspace{0.5cm} \\
    \hline
        configuration 1 & 3 & 1 & 1  & [0.10, 0.90]&31\\
        configuration 2 & 4 & 2 &  3 & [0.20, 0.80]& 241\\
        configuration 3 & 5 & 1 &  1 & [0.20, 0.80]&51\\
        configuration 4 & 5 & 1 &  2 & [0.15, 0.85]& 101\\
        configuration 5 & 10 & 1 &  3 & [0.10, 0.90]&301\\
        configuration 6 & 4 & 1 &  4 & [0.25, 0.75]& 161\\
    \hline
    \end{tabular}}
    \caption{\label{table:configurations_ridge_regression}Table of all configurations $V$, $G$, $r$ and [$a$, $b$] and their labels used in the analysis of the influence of the regression parameter on the forecasting ability.}
\end{table}

\clearpage
\newpage
\section{\label{sec:best_performer_info}Supplementary Note 2: Hyperparameterspace and best performing hyperprameter configurations}
The hyperparameterspace that is examined by Bayesian hyperparameter optimization is displayed in Table \ref{table: hyperparameters} and we list the best performing  hyperparameter sets that are used to obtain the results presented in main part of the paper, for each chaotic system in Table \ref{table:hyperparameter_choices}.
\begin{table}[h!]
\begin{center}
\resizebox{0.8\linewidth}{!}{
\begin{tabular}{ |l|l| }
 \hline
Parameter & Parameter range\\
 \hline
Number of evolutions: & $V \in\ \{1,2,...,15\}$\\
Number of reservoirs: & $r \in\ \{1,2,3\} $\\
Regression parameter: & $ 10^{-20} < \beta<10^{3}$\\
Readout polynominal: & $G \in \{1,2,3,4\}$ \\ 
Encoding interval: & $\begin{tiny} [a,b] \in \{[0.05,0.95],[0.10,0.90],[0.15,0.85],\hdots,[0.40,0.60],[0.45,0.55]\}\end{tiny}$\\
 \hline
\end{tabular}}
\caption{\label{table: hyperparameters}Hyperparameter space that is examined by the Bayesian hyperparameter search.}
\end{center}
\end{table}\\
\begin{table}[h]
    \centering
    \resizebox{0.8\linewidth}{!}{
    \begin{tabular}{|l|c|c|c|c|c|}
    \hline
    system \hspace{1.5cm}& \hspace{0.5cm} $V$ \hspace{0.5cm} & \hspace{0.5cm} $r$  \hspace{0.5cm} & $\beta$ & \hspace{0.5cm} $G$ \hspace{0.5cm} & \hspace{0.5cm} [$a$,$b$] \hspace{0.5cm} \\
    \hline
        Lorenz-63& 9 & 3 & 1.41$\cdot 10^{-12}$ & 3 & [0.15,0.85]\\
        Chen& 8 & 3 & 1.09$\cdot 10^{-12}$ & 3 & [0.30,0.70] \\
        Chua & 14 & 3 & 0.000269 & 2 & [0.10,0.90] \\
        Halvorsen & 8 & 3 & 1.41$\cdot 10^{-12}$ & 3 & [0.20,0.80]\\
        Rössler& 9 & 3 & 2.10$\cdot 10^{-12}$ & 3 & [0.20,0.80]\\
        Rucklidge& 7 & 3 & 1.25$\cdot 10^{-12}$ & 4 & [0.15,0.85]\\
        Thomas& 15 & 3 & 1.89 $\cdot 10^{-10}$& 4 & [0.05,0.95] \\
        WINDMI& 10 & 3 & 9.13$\cdot 10^{-12}$ & 4 & [0.05,0.95]\\
    \hline
    \end{tabular}}
    \caption{\label{table:hyperparameter_choices} Best performing Hyperparameter configuration for each chaotic system obtained by maximizing the forecast horizon with Bayesian hyperparameter optimization.}
\end{table}
\newpage
\section{\label{sec:unitary_operator_investigation}Supplementary Note 3: Investigation diverging trajectories (WINDMI)}
\begin{figure}[h!]
    \begin{center}
    \includegraphics[width=0.95\linewidth]{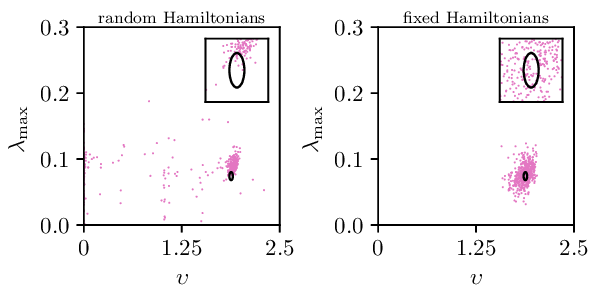}
    \end{center}
    \caption{\label{fig:unitary_fixed} Predicted climate for each of the 500 forecasted trajectories of the WINDMI system as presented in FIG. 2 (random Hamiltonians) in the main text and the same plot with fixed (well performing Hamiltonians) for the same hyperparameter configuration (see Table \ref{table:hyperparameter_choices}).}
\end{figure}

\begin{figure}[h!]
    \begin{center}
    \includegraphics[width=0.95\linewidth]{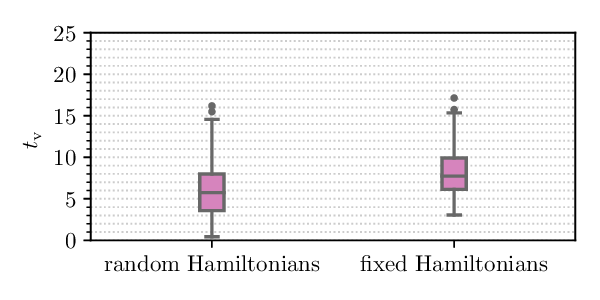}
    \end{center}
    \caption{\label{fig:unitary_fixed_2}Forecast horizon for randomly drawn Hamiltonians and fixed (well performing) Hamiltonians of the WINDMI system forecasting (500 trajectories each) with the hyperparameter configuration denoted in Table \ref{table:hyperparameter_choices}.}
\end{figure}

We do not optimize the unitary evolution, i.e., the quantum reservoir, due to the reasons described in the main text. The type of unitary evolution (reservoir) and the parameter ranges are not changed in the optimization process. For a more complete optimization the pipeline introduced in this paper should also include the optimization of the unitary operators. This is clearly shown by the results of a small numerical experiment forecasting the WINDMI system, where the model's climate prediction is the worst. In Fig. \ref{fig:unitary_fixed} and Fig. \ref{fig:unitary_fixed_2} again the results of the best performing hyperparameter combination are presented together with results from a slightly changed experimental setup. For one trajectory the model is trained with the hyperparameters presented in \ref{table:hyperparameter_choices} and therefore the unitary operators drawn from the uniform intervals until we find a model initialization with a forecast horizon that is larger than the mean forecast horizon of the best-performing model presented in the main part of the article. Afterward, 500 trajectories of the WINDMI system are trained and forecasted with exactly these unitary operators and hyperparameters. The long-term climate pattern of these results is presented in Fig. \ref{fig:unitary_fixed}. The two plots clearly show that the unitary operator ("network connections") has a huge impact on the prediction performance. The resulting mean forecast horizon is $8.1 \pm 2.6$ Lyapunov times. Hence, the short-term prediction quality is improved. The short-term prediction ability is further analyzed in Fig. \ref{fig:unitary_fixed_2}. It is clear that using a "well-performing" combination of unitary operators leads to the disappearance of the predicted trajectories that deviate after very short time scales from the real continuation of the time series, and also results in accurate climate prediction. 
\newpage
\section{\label{sec:chaos}Supplementary Note 4: Chaotic systems}
Three-dimensional chaotic systems described by a flow $F$ acting on the current state $\mathrm{\mathbf{u}}(t)$ of the system
\begin{equation}
\label{flow_equation}
\mathrm{\dot{\mathbf{u}}}(t)=F(\mathrm{\mathbf{u}}(t)).
\end{equation}
are used to display the potential QRC has in forecasting of nonlinear dynamical systems. The trajectories ($\mathrm{\mathbf{u}}(t)=\{\mathrm{\mathbf{u}}(t_{0}),\mathrm{\mathbf{u}}(t_{0}+\Delta t),\cdots \}$) are obtained by numerically solving Eq. \ref{flow_equation} using the fourth-order Runge-Kutta method (RK4) \cite{Runge1895, Kutta, sprott2003chaos}. In this study, eight three-dimensional prototypical chaotic systems with different system characteristics are tested.
The differential equations describing the systems, the selected system parameter choices and details about the generation of the datasets are defined in the following paragraph.\\

The discrete time series used in our work are obtained employing the RK4 method with the starting point, $\Delta t$, and the system parameters chosen as specified below, to obtain 25001000 consecutive steps of the attractor. We throw away the first 1000 steps and split the remaining time series into 1000 different trajectories each consisting of 25000 steps. The system defining equations are listed in the following and the defining parameters are denoted in Table \ref{table:system parameters}.
\begin{table*}[htb]
    \centering
    \resizebox{\linewidth}{!}{
    \begin{tabular}{|l|c|c|c|}
    \hline
    system & system parameters &  $\Delta t$ &\hspace{0.2cm} initial state \hspace{0.2cm} \\
    \hline
        Lorenz-63& $\rho$ = 28, $\sigma$ = 10, $\beta$ = 8/3 & \hspace{0.2cm} 0.02 \hspace{0.2cm} & [0, $-0.01$, 9]\\
        Chen & $a$ = 35, $b$ = 3, $c$ = 28 & 0.02 &[$-$10, 0, 37]\\
        Chua & \hspace{1cm} $\alpha$ = 9, $\beta$ = 100/7, a = 8/7, b = 5/7 \hspace{1cm} &  0.1 & [0, 0, 0.6]\\
        Halvorsen \hspace{1cm}& $a$ = 1.27 & 0.05 & [$-$5, 0, 0]\\
        Roessler & $a = b = 0.2, c = 5.7$ & 0.1 & [$-$9, 0, 0]\\
        Rucklidge& $\kappa$ = 2.0, $\lambda$ = 6.7 & 0.1 & [1, 0, 4.5]\\
        Thomas& $b = 0.18 $& 0.3 & [0.1, 0, 0]\\
        WINDMI& $a = 0.7, b = 2.5$ &  0.2 &[0, 0.8, 0]\\
    \hline
    \end{tabular}}
    \caption{\label{table:system parameters}System parameters, $\Delta t$ and initial state for all eight chaotic systems that are forecasted in this work. The system parameters and the initial states are taken from \cite{sprott2003chaos}.}
\end{table*}
\newpage
\twocolumngrid
\textbf{Lorenz-63 :}\\
\vspace{-1cm}
\begin{gather} 
\Dot{x}=\sigma(y-x) \notag\\
\Dot{y}=x(\rho - z) - y \\
\Dot{z}=xy - \beta z \notag
\end{gather}

\textbf{Chen's system :}\\
\vspace{-1cm}
\begin{gather} 
\Dot{x} = a(y - x) \notag\\
\Dot{y} = (c-a)x-xz+cy\\
\Dot{z} = xy-bz \notag
\end{gather}
\textbf{Chua's circuit :}\\
\vspace{-1cm}
\begin{small}
\begin{gather} 
\Dot{x} = \alpha[y-x+bx+0.5(a-b)(\vert x + 1 \vert-\vert x - 1 \vert)] \notag\\
\Dot{y} = x-y+z\\
\Dot{z} = -\beta y  \notag
\end{gather}
\end{small}
\textbf{Halvorsen attractor :}\\
\vspace{-1cm}
\begin{gather} 
\Dot{x} = -ax-4y-4z-y^{2}\notag\\
\Dot{y} = -ay-4z-4x-z^{2}  \\
\Dot{z} = -az-4x-4y-x^{2}\notag
\end{gather}
\textbf{Rössler attractor :}\\
\vspace{-1cm}
\begin{gather} 
\Dot{x} = -y-z \notag\\
\Dot{y} = x+ay  \\
\Dot{z} = b+z(x-c)  \notag
\end{gather}
\textbf{Rucklidge attractor :}\\
\vspace{-1cm}
\begin{gather} 
\Dot{x} = -\kappa x+\lambda y-yz  \notag\\
\Dot{y} = x  \\
\Dot{z} = -z+y^{2} \notag
\end{gather}
\textbf{Thomas attractor :}\\
\vspace{-1cm}
\begin{gather} 
\Dot{x} = -bx+\mathrm{sin}(y)  \notag\\
\Dot{y} = -by+\mathrm{sin}(z)  \\
\Dot{z} = -bz+\mathrm{sin}(x)  \notag
\end{gather}
\textbf{WINDMI attractor :}\\
\vspace{-1cm}
\begin{gather} 
\Dot{x} = y  \notag\\
\Dot{y} = z  \\
\Dot{z} = -az-y+b-\mathrm{exp}(x)\notag
\end{gather}
\let\cleardoublepage\clearpage

\onecolumngrid
\section{\label{sec:chaos}Supplementary Note 5: Noisy quantum systems}
\begin{figure*}[h!]
    \begin{center}
    \includegraphics[width=\textwidth]{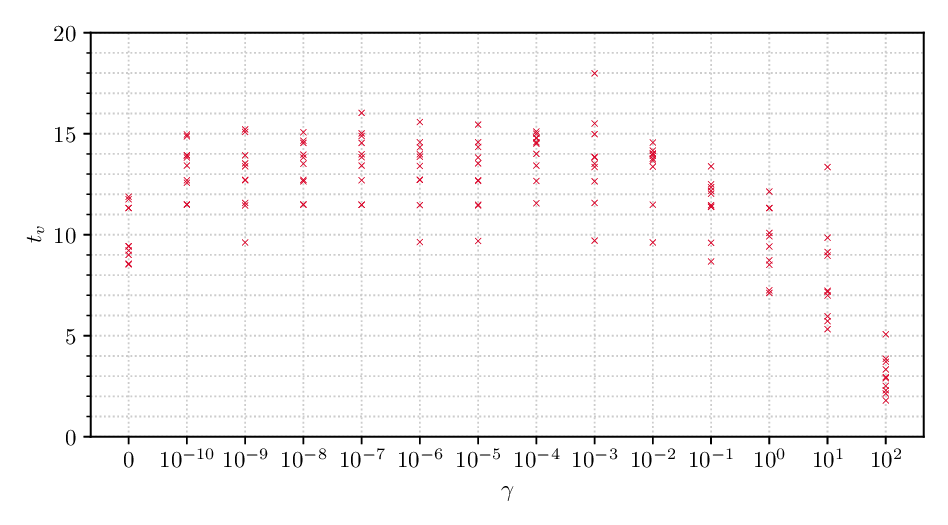}
    \end{center}
    \caption{\label{fig:noise}Forecast horizon of the prediction of the Lorenz system obtained by noisy quantum reservoirs with $[a,b]$=[0.15,0.85], $\beta=1.41 \times 10^{-12}$, $G=4$ and $V=9$ for various $\gamma$.}
\end{figure*}

The results of this work were obtained with the simulation of perfect quantum systems. We simulated noiseless unitary evolution of the quantum systems. For real world application this is not a realistic assumption because real quantum systems are not noiseless. One important step towards real world application is to investigate the effect of noise on QRC with the goal of finding unitary evolutions that can be implemented on real hardware, considering hardware limitations. This goes beyond the scope of the paper. Nevertheless we want to mention that there are first results showing that QRC seems to be quite resistant to noise and even benefit from a certain amount of noise in some cases \cite{Domingo2023_noise, Fujii_2017}. This gives hope that applications on NISQ era devices might be possible. Furthermore, some initial results are presented that showcase that the described QRC framework using minimal amount (namely four) of qubits show a similar resistance against noise in the quantum reservoirs. The results are obtained by going from unitary evolution of the quantum reservoirs to evolutions obtained by solving the Lindblad equation numerically using the python package qutip \cite{JOHANSSON20131234}. As Lindblad operators, we consider for each qubit $L_{i}=\sqrt{\gamma}\sigma^{i}_{z}$. This describes a system with dephasing. We forecast the Lorenz system using one reservoir ($r$=1) for different noise levels (controlled by $\gamma$). For each noise level, we forecast $N_{\mathrm{stat}}=10$ trajectories for $N_{\mathrm{pred}}$=2000 steps. The forecasting results for each of these realizations are plotted in Fig. \ref{fig:noise}. One can observe that for a large range of $\gamma$, the prediction remains accurate on longer timescales compared to the noiseless case ($\gamma=0$). As the noise increases further, the forecast horizon shortens, but the prediction does not completely break down.

\newpage

\clearpage
\providecommand{\noopsort}[1]{}\providecommand{\singleletter}[1]{#1}%
%